\documentclass[
 reprint,
 amsmath, amssymb,
 aps,
prb,
]{revtex4-2}

\usepackage{graphicx} 
\usepackage{dcolumn}
\usepackage{bm}
\usepackage{cancel}
\usepackage{multirow}
\usepackage{here}
\usepackage{enumerate}

\usepackage{braket}
\usepackage[usenames, dvipsnames]{xcolor} 
\usepackage{amsmath}
\usepackage{comment}

\begin{document}

\preprint{APS/123-QED}

\title{Systematic analysis method for nonlinear response tensors}

\author{Rikuto Oiwa$^{1}$}
\author{Hiroaki Kusunose$^{1}$}%
\affiliation{%
  $^{1}$Department of Physics, Meiji University, Kawasaki 214-8571, Japan
}%

\date{\today}

\begin{abstract}
  We propose a systematic analysis method for identifying essential parameters in various linear and nonlinear response tensors without which they vanish.
  By using the Keldysh formalism and the Chebyshev polynomial expansion method, the response tensors are decomposed into the model-independent and dependent parts, in which the latter is utilized to extract the essential parameters.
  An application of the method is demonstrated by analyzing the nonlinear Hall effect in the ferroelectric SnTe monolayer for example.
  It is shown that in this example the second-neighbor hopping is essential for the nonlinear Hall effect whereas the spin-orbit coupling is unnecessary.
  Moreover, by analyzing terms contributing to the essential parameters in the lowest order, the appearance of the nonlinear Hall effect can be interpreted by the subsequent two processes: the orbital magneto-current effect and the linear anomalous Hall effect by the induced orbital magnetization.
  In this way, the present method provides a microscopic picture of responses.
  By combining with computational analysis, it stimulates further discoveries of anomalous responses by filling in a missing link among hidden degrees of freedom in a wide variety of materials.
\end{abstract}

\maketitle

\section{Introduction}
\label{sec:intro}

A variety of linear and nonlinear responses in cooperation with a magnetic ordering has been paid much attention recently in view of the so-called Berry-curvature mechanism.
For example, the anomalous Hall effect (AHE)~\cite{Nagaosa_2010, Xiao_2010, Gradhand_2012}, Kerr effect~\cite{Argyres_1955, NAGAMIYA_1982, Feng_2015, Higo_2018}, Nernst effect~\cite{Luttinger_1964, Xiao_2006, Ikhlas_2017, Destraz_2020}, magnetoelectric effect~\cite{Schmid_2008, Hayami_2014, Saito_2018, Khanh_2016, Khanh_2017, Yanagi_201801, Yanagi_201802, Shinozaki_2020}, magnetopiezoelectric effect~\cite{HW_YY_MPE_2017, Shiomi_2019, Shiomi_2020}, and nonreciprocal transports in multiferroic materials~\cite{Tokura2018} have been investigated.
In particular, the nontrivial AHE under the collinear~\cite{Crystal_trsb_2020, li_2019, feng_2021, Ket_2020, SH_2021} and non-collinear~\cite{Chen_2014, K_bler_2014, Nakatsuji_mn3sn_2015, Nakatsuji_mn3ge_2016, Nayak_mn3ge_2016, Yang_2017, Akiba_2020} antiferromagnetic (AFM) orderings has been studied extensively.

In addition to the above linear responses, the fascinating nonlinear responses such as the nonlinear optoelectronic transport~\cite{Matsyshyn_2019, HW_nonlinear_cond_2020, HW_nonlinear_optoele_2021}, nonlinear Nernst effect~\cite{Yu_2019, Zeng_2019, Zeng_2020, Yu_2021}, and nonlinear magnetoelectric effect~\cite{Kimura_2003, Cao_2017} have also been elucidated.
For instance, the nonlinear Hall effect (NLHE) has been studied in ferroelectriclike metals~\cite{Xiao_2020}, thin films of a Weyl semimetal~\cite{Morimoto_2016, Xu_wte2_2018, Du_wte2_2018, Ma_wte2_2019, Kang2019, Wang2019, Xiao_wte2_2020}, monolayer strained MoS$_{2}$~\cite{Son_2019}, an organic Dirac fermion system~\cite{Osada_2020}, elemental Te~\cite{Tsirkin_Te_fp2_2018, snte_Wangeaav_2019, snte_kim_2019}, and so on.

The NLHE is often characterized by the Berry curvature dipole (BCD)~\cite{Sodemann_bcd_2015} that is a measure of the dipole structure of the Berry curvature (BC) in the momentum space.
The realization of NLHE in the ferroelectric atomic-thick SnTe has been proposed by the {\it ab initio} calculation and model analysis~\cite{snte_Chang_2016}, where it is indicated that a coupling between the ferroelectric order parameter and BC is crucial.

In the series of study for the gigantic AHE observed in the non-collinear AFM Mn$_{3}$Sn~\cite{Nakatsuji_mn3sn_2015}, a methodology combining the {\it ab initio} calculation~\cite{Wang_wannier_2006} with the symmetry-adopted multipole theory~\cite{SH_HK_2018, SH_MY_YY_HK_mul_2018, HW_YY_Mul_2018, HK_RO_SH_comp_mul_2020} has been developed~\cite{MTS_cmul_ahe_2018} as well as the Berry-curvature mechanism~\cite{Anderson_1955, Ohgushi_2000, Taguchi_2001, Zhang_2020}.
Meanwhile, the recent study indicates the relevance of the anisotropic magnetic dipole observed in the x-ray magneto-circular dichroism (XMCD)~\cite{Carra_1993, Stohr_PRL_1995, Stohr_JELEC_1995, Crocombette_1996, Yamasaki_2020} to the AHE in several antiferromagnets~\cite{SH_2021,HK_RO_SH_comp_mul_2020}.

As mentioned above, linear and nonlinear responses have mainly been analyzed by the {\it ab initio} calculations, the topological structure of electronic bands, and the group theoretical argument with electronic multipoles.
Although these approaches are successful in several aspects, it is still insufficient to understand the microscopic picture of the responses, e.g., which model parameters are essential without which a response vanishes, especially in the nonlinear responses.
Once the essential parameters are identified, they provide a microscopic picture of the response, i.e., minimal couplings between the model parameters, such as electrons hopping, spin-orbit coupling (SOC), etc., and electric and/or magnetic order parameters.

In this paper, we propose a systematic analysis method for linear and nonlinear response tensors, which enables us to extract essential model parameters of a given theoretical model.
Our method is based on the Keldysh formalism~\cite{Keldysh_1964, jishi_2013, Jo_o_2019} in order to treat the nonlinear response tensors systematically and avoid the analytic continuation procedure in the Matsubara formalism.
Following the Chebyshev polynomial expansion method~\cite{Alexander_KPM_2006, Amrendra_2002, Braun_2014, Ferreira_2015, Jo_o_2019}, the response tensors are decomposed into the model-independent and dependent parts: the latter is expressed as a power series of the Hamiltonian matrix and operators of a response.
By analyzing the resultant low-order contributions in the model-dependent part analytically, we can identify essential parameters and deduce a microscopic picture of the response as a minimal coupling between the degrees of freedom associated with the essential parameters.
For a specific example, we discuss the NLHE of the ferroelectric monolayer SnTe, and demonstrate how we can identify essential parameters and deduce a microscopic picture of the response.
Identifying the essential parameters for various responses could lead to a deeper understanding of the response and a useful guideline for efficient design of future functional materials.
The organization of the paper is given in the next outline section.

\begin{figure*}[t!]
  \begin{center}
    \includegraphics[width=15cm]{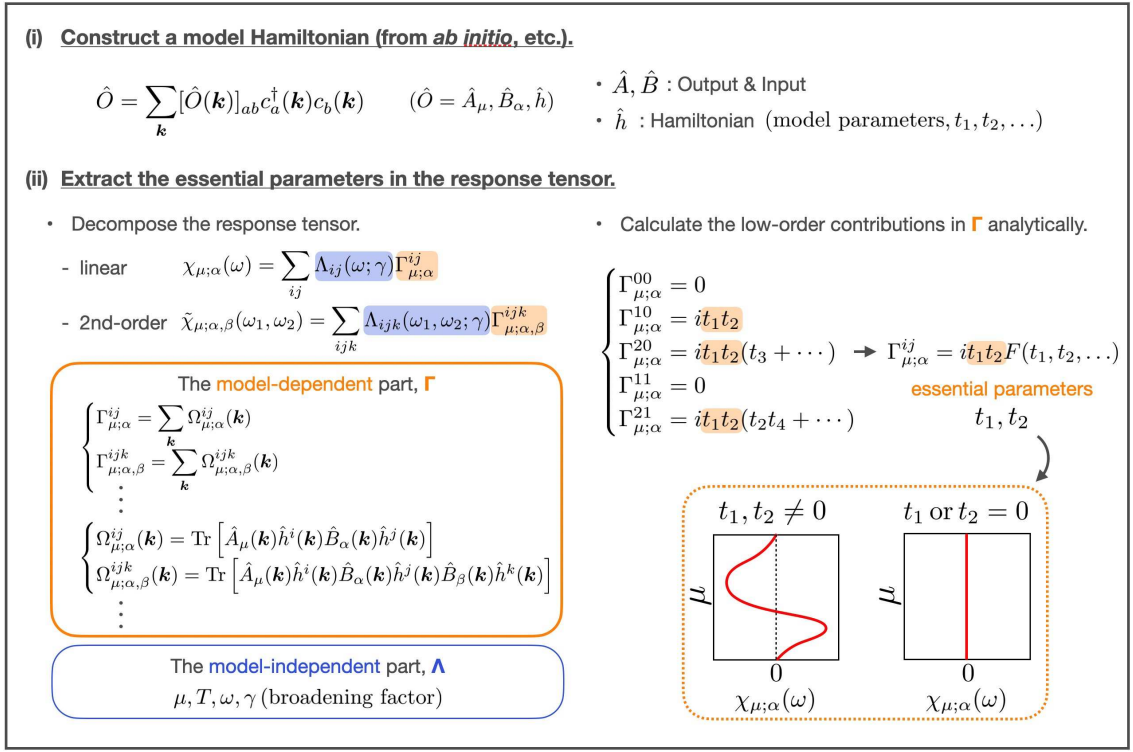}
  \end{center}
  \caption{
    Outline of the systematic analysis method for identifying essential parameters in the response tensors.
    By evaluating the model-dependent $\Gamma$ part, we obtain the essential parameters.
  }
  \label{fig_outline}
\end{figure*}

\section{Outline}
\label{sec:outline}

In this section, we give the organization of the paper and the outline of the systematic analysis method to extract the essential parameters in linear and nonlinear responses for a given model Hamiltonian.
The main points of the method are summarized in Fig.~\ref{fig_outline}.

The present method consists of the following two parts:

\begin{enumerate}[(i)]
  \item Construct a one-body model Hamiltonian $\hat{h}$ and set the output (response) and input (external fields) operators $\hat{A}$, $\hat{B}$ in the response.
        The Hamiltonian $\hat{h}$ involves the model parameters $\{ t_{1}, t_{2}, \ldots \}$ including the mean-field terms of an electronic ordering.

  \item Extract the essential parameters in the response tensor.
        In general, we can decompose the expression of response tensors into the model-independent ($\Lambda$) and dependent ($\Gamma$) parts.
        Since all the model parameters and symmetry information of the response tensor, i.e., $\hat{A}$, $\hat{B}$, and $\hat{h}$, appear only in the latter part $\Gamma$, we can extract the essential parameters by evaluating this part.

        In the momentum-space representation, $\Gamma$ is expressed as the sum of $\Omega (\bm{k})$, which consists of the trace of the products of $\hat{A}$, $\hat{B}$, and powers of $\hat{h}$.
        By evaluating the low-order contributions of $\Gamma$ analytically, the resultant expressions have common proportional factors as shown in the bottom-right part of Fig.~\ref{fig_outline}.
        Then, the overall contributions are proportional to the common factors, which are nothing but the essential parameters as the response tensor vanishes if one of them is set to zero.

        By the systematic analysis of the higher-order terms, and the comparison with the numerical evaluation of the response if necessary, we can confirm the relevance of the essential parameters.
\end{enumerate}

For the purpose of extracting the essential parameters, we mainly focus on the analysis of the model-dependent $\Gamma$ part in this paper.
However, the whole expression in the decomposed form is useful even for numerical calculation~\cite{Jo_o_2019}.
Since the (nonlinear) Kubo formula often involves the higher-order $\bm{k}$ derivatives of the energy dispersion, the direct momentum summation in the Kubo formula results in a serious reduction of numerical accuracy.
The present expressions do not suffer from such difficulty, although considerably high-order terms and the explicit expression of $\Lambda$ part are necessary to obtain the convergence of the expansion.

In order to make the present paper self-contained, the paper is organized as follows.
In Sect.~\ref{sec:formulation}, we begin with a brief review of the Keldysh formalism and the Chebyshev polynomial expansion method.
Following the work by S. M. Jo{\~{a}}o and his coworkers~\cite{Jo_o_2019}, we show the basis-independent form of the response tensors and its power series of the given Hamiltonian.
Then, we discuss the symmetry of $\Lambda$ and $\Gamma$ which is useful to reduce the relevant terms, especially for the static response $\omega=0$.
In Sect.~\ref{sec:params_extraction}, we give the explicit formula in the momentum space in the thermal average, and linear and nonlinear responses, which corresponds to the specific procedures of (i) and (ii).
In Sect.~\ref{sec:cond}, we exhibit the expressions specific for the linear and nonlinear electrical conductivities, as they require special care in the velocity operator that depends on the input electric field.
Then, for a practical example, we discuss the essential parameters and the microscopic picture of the NLHE in the ferroelectric SnTe monolayer in Sect.~\ref{sec:snte}.
The final section summarizes the paper.
There are three appendices and supplemental material.
In Appendix~\ref{sec:kel}, we give a brief introduction of the Keldysh formalism.
In Appendix~\ref{sec:nlc}, the derivation of the $n$-th order conductivity in the velocity gauge is given.
The derivation of the linear conductivity {\it in the velocity gague} is given in Appendix~\ref{sec:lcon_vel}
The derivation and formula of the other responses such as the electric-field/current induced ones are given in the supplemental material.

\section{General Expression}
\label{sec:formulation}

In this section, we briefly review the Chebyshev polynomial expansion method~\cite{Alexander_KPM_2006, Amrendra_2002, Braun_2014, Ferreira_2015, Jo_o_2019} in the Keldysh formalism~\cite{Keldysh_1964, jishi_2013, Jo_o_2019} in order to treat linear and nonlinear response tensors analytically in a self-contained manner.
Following the previous work~\cite{Jo_o_2019}, the general expressions of the response tensors are given in terms of the Keldysh Green's functions in Sect.~\ref{sec:res_tensor_keldysh}, and then they are converted to the separable form by means of the Chebyshev polynomial expansion~\cite{Jo_o_2019} in Sect.~\ref{sec:cp_ex}.
The separable form is further rearranged based on the symmetry property in Sect.~\ref{sec:cp_slg}: The resultant form is the most suitable for analyzing the essential parameters in the responses.

In this paper, we restrict ourselves to electronic systems expressed by a one-body Hamiltonian including those given by the {\it ab initio} and mean-field calculations.

\subsection{Response Tensor}
\label{sec:res_tensor_keldysh}

Let us consider a Hamiltonian given by
\begin{align}
  \mathcal{H}(t) & = \mathcal{H}_{0} + \mathcal{H}_{\rm ext} (t),
  \label{eq_ham}
\end{align}
where $\mathcal{H}_{0}$ is the one-body Hamiltonian of our system, and $\mathcal{H}_{\rm ext} (t)$ is the time-dependent perturbation for arbitrary external fields.
They are explicitly given by
\begin{align}
   & \mathcal{H}_{0}            = \sum_{ij} [\hat{H}_{0}]_{ij} c_{i}^{\dagger} c_{j}^{},
  \label{eq_H0_normal}                                                                                                       \\
   & \mathcal{H}_{\rm ext} (t)  = \sum_{\alpha} F_{\alpha}(t) \sum_{ij} [\hat{B}_{\alpha}]_{ij}(t) c_{i}^{\dagger} c_{j}^{},
  \label{eq_Hext_pert}
\end{align}
where $c^{(\dagger)}_{i}$ is the annihilation (creation) operator of an arbitrary state $i$, such as a set of the momentum, spin, orbital, and sublattice.
$F_{\alpha}(t)$ denotes the $\alpha$-th component of the time-dependent generalized force coupled with an hermitian electron operator $\hat{B}_{\alpha}$ with or without explicit $t$ dependence.
The repeated indices, e.g., $\alpha$, $i$, $j$, are implicitly summed over hereafter.

The ensemble average of an arbitrary hermitian operator labelled by $\mu$ is given by
\begin{align}
  A_{\mu}(t) \equiv \braket{\hat{A}_{\mu}^{\rm H}(t)} = -\, \mathrm{Tr} \left[ \hat{A}_{\mu}(t) i \hat{G}^{<}(t, t) \right],
  \label{eq_At_av}
\end{align}
where the superscript H denotes the Heisenberg representation, $\hat{O}^{\rm H}(t) = e^{i\mathcal{H}t/\hbar} \hat{O} e^{-i\mathcal{H}t/\hbar}$ and $\hat{G}^{<}$ represents the matrix form of the lesser Green's function, whose $(i,j)$-component is defined as
\begin{align}
   & i G^{<}_{ij} (t, t') = - \braket{c_{j}^{\dagger, {\rm H}}(t') c_{i}^{{\rm H}}(t) }_{0}.
\end{align}
Here, $\braket{\cdots}_{0}$ is the ensemble average with respect to $\mathcal{H}_{0}$.

Using the perturbation expansion of $\hat{G}^{<}$ against the external field in the Keldysh formalism (see Appendix~\ref{sec:kel} in detail), the $n$-th order contribution of Eq.~(\ref{eq_At_av}) in the frequency domain is given by
\begin{align}
  A_{\mu}^{(n)}(\omega)
  =
  -\int_{\omega_{a}} \mathrm{Tr} \left[\hat{A}_{\mu}(\omega_{a}) i \hat{G}^{< (n)}(\omega - \omega_{a})\right],
  \label{eq_Ao_av_n}
\end{align}
where $\int_{\omega} \equiv \int_{-\infty}^{\infty} d\omega/2\pi$ for notational simplicity, and $\hat{G}^{< (n)}$ is the $n$-th order lesser Green's function (see Appendix~\ref{sec:kel} in detail).
The explicit expressions of $A_{\mu}^{(n)}(\omega)$ up to $n = 2$ are given by
\begin{widetext}
  \begin{align}
        &
    A_{\mu}^{(0)}(\omega)
    =
    -i \int_{\omega_{c_{1}}}
    \mathrm{Tr}
    \left[
      \hat{A}_{\mu}(\omega)
      \hat{\mathcal{G}}^{<}\left(\omega_{c_{1}}\right)
      \right],
    \label{eq_Ao_0}
    \\
        &
    A_{\mu}^{(1)}(\omega)
    =
    -i\frac{(2\pi)^{2}}{\hbar}
    \int_{\omega_{a},\omega_{1},\omega_{c_{1}\cdots c_{3}}}
    F_{\alpha}(\omega_{1})
    \delta(\omega_{c_{1}} - \omega_{c_{2}} - \omega_{c_{3}})
    \delta(\omega - \omega_{a} + \omega_{c_{3}} - \omega_{c_{1}})
    \cr & \hspace{4cm}
    \times
    \mathrm{Tr}\left[
      \hat{A}_{\mu}(\omega_{a})
      \left\{
      \hat{\mathcal{G}}^{<}\left(\omega_{c_{1}}\right)
      \hat{B}_{\alpha}(\omega_{c_{2}} - \omega_{1})
      \hat{\mathcal{G}}^{\rm A}\left(\omega_{c_{3}}\right)
      +
      \hat{\mathcal{G}}^{\rm R}\left(\omega_{c_{1}}\right)
      \hat{B}_{\alpha}(\omega_{c_{2}} - \omega_{1})
      \hat{\mathcal{G}}^{<}\left(\omega_{c_{3}}\right)
      \right\}
      \right],
    \label{eq_Ao_1}
  \end{align}
  and
  \begin{align}
                       &
    A_{\mu}^{(2)}(\omega)
    =
    -i \frac{(2\pi)^{3}}{\hbar^{2}}
    \int_{\omega_{a},\omega_{1},\omega_{2},\omega_{c_{1}\cdots c_{5}}}
    F_{\alpha}(\omega_{1}) F_{\beta}(\omega_{2})
    \delta(\omega_{c_{1}} - \omega_{c_{2}} - \omega_{c_{3}})
    \delta(\omega_{c_{3}} - \omega_{c_{4}} - \omega_{c_{5}})
    \delta(\omega - \omega_{a} + \omega_{c_{5}} - \omega_{c_{1}})
    \cr                & \hspace{5cm}
    \times
    \mathrm{Tr} \left[
      \hat{A}_{\mu}(\omega_{a}) \left\{
      \hat{\mathcal{G}}^{<}\left(\omega_{c_{1}}\right)
      \hat{B}_{\alpha}(\omega_{c_{2}} - \omega_{1})
      \hat{\mathcal{G}}^{\rm A}\left(\omega_{c_{3}}\right)
      \hat{B}_{\beta}(\omega_{c_{4}} - \omega_{2})
      \hat{\mathcal{G}}^{\rm A}\left(\omega_{c_{5}}\right)
    \right.\right. \cr & \hspace{6cm}
      +
      \hat{\mathcal{G}}^{\rm R}\left(\omega_{c_{1}}\right)
      \hat{B}_{\alpha}(\omega_{c_{2}} - \omega_{1})
      \hat{\mathcal{G}}^{<}\left(\omega_{c_{3}}\right)
      \hat{B}_{\beta}(\omega_{c_{4}} - \omega_{2})
      \hat{\mathcal{G}}^{\rm A}\left(\omega_{c_{5}}\right)
    \cr                & \hspace{7.2cm} \left.\left.
      +
      \hat{\mathcal{G}}^{\rm R}\left(\omega_{c_{1}}\right)
      \hat{B}_{\alpha}(\omega_{c_{2}} - \omega_{1})
      \hat{\mathcal{G}}^{\rm R}\left(\omega_{c_{3}}\right)
      \hat{B}_{\beta}(\omega_{c_{4}} - \omega_{2})
      \hat{\mathcal{G}}^{<}\left(\omega_{c_{5}}\right)
      \right\}\right]
    \label{eq_Ao_2},
  \end{align}
\end{widetext}
where, $\hat{\mathcal{G}}^{\rm R}, \hat{\mathcal{G}}^{\rm A},$ and $\hat{\mathcal{G}}^{<}$ are the unperturbed retarded, advanced, and lesser Green's functions, respectively.
Their matrix forms are explicitly given by
\begin{align}
   & \hat{\mathcal{G}}^{<}(\omega)
  =
  2\pi i \hbar\, f(\hbar \omega) \delta(\hbar \omega - \hat{H}_{0}),
  \label{eq_g_lesser}                                                                                 \\
   & \hat{\mathcal{G}}^{\zeta}(\omega) = \frac{\hbar}{\hbar (\omega+ \zeta i \gamma) - \hat{H}_{0} },
  \label{eq_G_RA}
\end{align}
where $\gamma > 0$ is a broadening factor proportional to the inverse of the phenomenological relaxation time $\tau = \gamma^{-1}$, and $\zeta = + (\rm R)$ and $\zeta = - (\rm A)$ correspond to the retarded and advanced Green's functions, respectively.
$f (E) = (e^{\beta ( E - \mu )} + 1)^{-1}$ is the Fermi-Dirac distribution function, where $\beta$ and $\mu$ denote the inverse temperature and the chemical potential.

From the above expressions, we define the $n$-th order response tensor $\chi_{\mu; \alpha_{1}, \ldots, \alpha_{n}} \left(\omega_{1}, \ldots, \omega_{n}\right)$ as
\begin{multline}
  A_{\mu}^{(n)}(\omega)
  = 2\pi \left(\prod_{i = 1}^{n} \int_{\omega_{i}}
  F_{\alpha_{i}}\left(\omega_{i}\right)\right) \delta(\omega - \omega_{[n]})
  \\
  \times \chi_{\mu; \alpha_{1}, \ldots, \alpha_{n}} \left(\omega_{1}, \ldots, \omega_{n}\right),
  \label{eq_Ao_n_ex}
\end{multline}
where $\omega_{[n]} = \sum_{i=1}^{n} \omega_{i}$.
In the subscript, $\mu$ represents the label of the response (output), while the others are those of the external fields (input).
By definition, $\chi$ is fully symmetric for any exchange of pair of external indices, and hence only the fully symmetric part of any exchange $(\alpha_{i}, \omega_{i})\leftrightarrow (\alpha_{j}, \omega_{j})$ contributes to the response, namely,
\begin{multline}
  \chi_{\mu; \alpha_{1}, \ldots, \alpha_{n}} \left(\omega_{1}, \ldots, \omega_{n}\right)
  \\
  = \frac{1}{n!} \sum_{\set{P}} \tilde{\chi}_{\mu; \alpha_{p_{1}}\ldots \alpha_{p_{n}}}(\omega_{p_{1}}, \ldots, \omega_{p_{n}}),
  \label{eq_Ao_n_ex_sym}
\end{multline}
where $\tilde{\chi}_{\mu; \alpha_{1}, \ldots, \alpha_{n}}\left(\omega_{1}, \ldots, \omega_{n}\right)$ is the non-symmetrized response tensor given below, and $\sum_{\set{P}}$ represents the sum over all permutations of $(1,2,\cdots,n)$.

\begin{figure}[t!]
  \begin{center}
    \includegraphics[width=9cm]{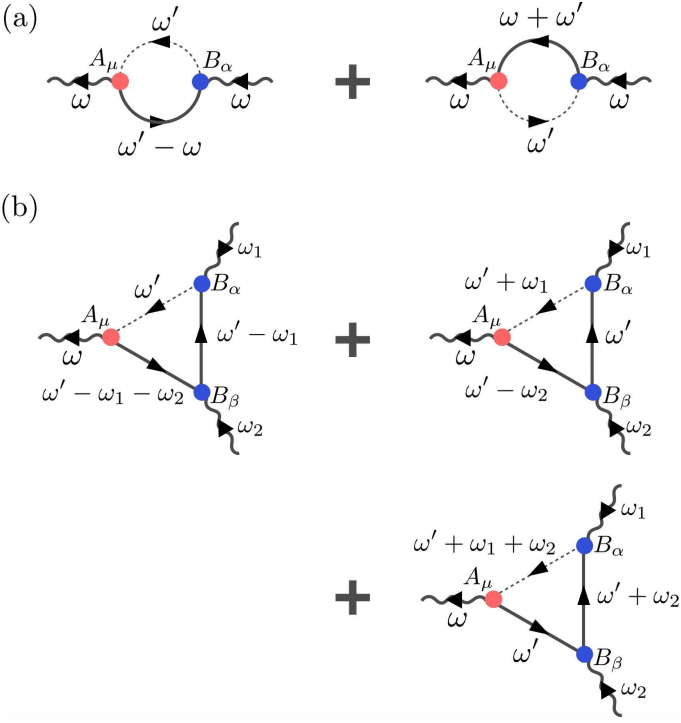}
  \end{center}
  \caption{
    Feynman diagrams for the (a) linear and (b) second-order response tensors.
    The incoming (outgoing) wavy line denotes the external field (response) with associated frequency.
    The blue (red) solid circle represents the input (output) operator.
    The dashed line in the closed diagram denotes the lesser Green's function.
    In the anti-clockwise closed loop starting at the outgoing vertex (red circle), the previous (subsequent) solid lines of the dashed line correspond to the advanced (retarded) Green's functions.
    The same rule can be applied to the higher-order terms.
  }
  \label{fig_X_diagram}
\end{figure}

Comparing Eq.~(\ref{eq_Ao_av_n}) with Eq.~(\ref{eq_Ao_n_ex}), the $n$-th order non-symmetrized nonlinear response tensor is obtained as
\begin{widetext}
  \begin{multline}
    \tilde{\chi}_{\mu; \alpha_{1}, \ldots, \alpha_{n}} \left(\omega_{1}, \ldots, \omega_{n}\right)
    =
    -i \frac{1}{\hbar^{n}} \int_{\omega_{c}} \sum_{\set{C}}
    \mathrm{Tr}\left[
    \hat{A}_{\mu}
    \left\{
    \hat{\mathcal{G}}^{C_{1}}(\omega_{c}) \hat{B}_{\alpha_{1}} \hat{\mathcal{G}}^{C_{2}}(\omega_{c} - \omega_{1}) \right.\right.\\
    \left.\left.\times \cdots \times
    \hat{\mathcal{G}}^{C_{n}}(\omega_{c} - \omega_{[n-1]})
    \hat{B}_{\alpha_{n}} \hat{\mathcal{G}}^{C_{n+1}}(\omega_{c} - \omega_{[n]})
    \right\}
    \right].
    \label{eq_chi_n}
  \end{multline}
  Here, the summation $\sum_{\set{C}}$ means that for given $k$, $(\hat{\mathcal{G}}^{C_{1}}, \ldots, \hat{\mathcal{G}}^{C_{k-1}})$, $\hat{\mathcal{G}}^{C_{k}}$, and $(\hat{\mathcal{G}}^{C_{k+1}}, \ldots, \hat{\mathcal{G}}^{C_{n+1}})$ are replaced with $\hat{\mathcal{G}}^{\rm R}$, $\hat{\mathcal{G}}^{<}$, and $\hat{\mathcal{G}}^{\rm A}$, respectively, where $k$ runs over from $1$ to $n+1$.

  In the case that both $\hat{A}_{\mu}$ and $\hat{B}_{\alpha}$ are time-independent operators, by replacing $\hat{\mathcal{G}}^{<}$ with Eq.~(\ref{eq_g_lesser}), the linear response tensor is expressed as
  \begin{align}
    \tilde{\chi}_{\mu; \alpha}(\omega)
    =
    \frac{1}{\hbar} \int_{E_{\rm min}}^{E_{\rm max}}  dE\, f(E)\,
    \mathrm{Tr}
    \left[
    \hat{A}_{\mu}
    \left\{ \delta(E - \hat{H}_{0}) \hat{B}_{\alpha} \hat{\mathcal{G}}^{{\rm A}}\left(E/\hbar - \omega\right)
    +
    \hat{\mathcal{G}}^{{\rm R}}\left(E/\hbar + \omega\right) \hat{B}_{\alpha} \delta(E - \hat{H}_{0})\right\}
    \right],
    \label{eq_X1_ab}
  \end{align}
  where we have introduced the finite energy range $[E_{\rm min},E_{\rm max}]$ of the spectrum of $\hat{H}_{0}$ because of the factor $\delta(E - \hat{H}_{0})$.
  Note that he linear term satisfies $\tilde{\chi}_{\mu; \alpha} = \chi_{\mu; \alpha}$ by definition.

  Similarly, the non-symmetrized second-order response tensor is given by
  \begin{multline}
    \tilde{\chi}_{\mu; \alpha, \beta}(\omega_{1}, \omega_{2})
    =
    \frac{1}{\hbar^{2}}
    \int_{E_{\rm min}}^{E_{\rm max}} dE\, f(E)\,
    \mathrm{Tr} \left[
      \hat{A}_{\mu} \left\{
      \delta(E - \hat{H}_{0}) \hat{B}_{\alpha} \hat{\mathcal{G}}^{\rm A} \left(E/\hbar - \omega_{1}\right) \hat{B}_{\beta} \hat{\mathcal{G}}^{\rm A}\left(E/\hbar - \omega_{1} - \omega_{2}\right)
      \right.\right.\\ \left.\left.
      +
      \hat{\mathcal{G}}^{\rm R}\left(E/\hbar + \omega_{1}\right) \hat{B}_{\alpha} \delta(E - \hat{H}_{0}) \hat{B}_{\beta} \hat{\mathcal{G}}^{\rm A}\left(E/\hbar - \omega_{2}\right)
      +
      \hat{\mathcal{G}}^{\rm R}\left(E/\hbar+ \omega_{1} + \omega_{2}\right) \hat{B}_{\alpha} \hat{\mathcal{G}}^{\rm R} \left(E/\hbar + \omega_{2}\right) \hat{B}_{\beta} \delta(E - \hat{H}_{0})
      \right\}\right]
    .
    \label{eq_X2_ab}
  \end{multline}
\end{widetext}
The Feynman diagrams corresponding to Eqs.~(\ref{eq_X1_ab}) and (\ref{eq_X2_ab}) are shown in Fig.~\ref{fig_X_diagram}.
The higher-order response tensors are systematically obtained by the same procedure.
Since Eqs.~(\ref{eq_X1_ab}) and (\ref{eq_X2_ab}) are independent of a choice of one-body basis, these expressions can be applied to systems irrespective of the presence of translational invariance~\cite{Jo_o_2019}.

When $\hat{A}$ or $\hat{B}$ has an implicit external-field dependencies, e.g., an electric current operator in the velocity gauge, we have to further expand $\hat{A}$ or $\hat{B}$ with respect to them in Eq.~(\ref{eq_Ao_av_n}), giving rise to the additional terms in Eqs.~(\ref{eq_X1_ab}) and (\ref{eq_X2_ab}).
The typical example will be shown in Sect.~\ref{sec:cond} for the electric conductivities.

\subsection{Chebyshev Polynomial Expansion}
\label{sec:cp_ex}
Here, the Chebyshev polynomial expansion method is introduced~\cite{Alexander_KPM_2006, Amrendra_2002, Braun_2014, Ferreira_2015, Jo_o_2019} to decompose the response tensors into the model independent ($\Lambda$) and dependent ($\Gamma$) parts.

In order to adopt the Chebyshev polynomial expansion, we first rescale all the energies in the range $\mathcal{I} = [-1+\delta, 1-\delta]$ with a positive small parameter $\delta < 1$ as
\begin{align}
   & \hat{H}_{0} \to \hat{h} = \frac{\hat{H}_{0} - E_{\rm c} \hat{I}}{W},
  \label{eq_H_rescaled}                                                   \\
   & E        \to \epsilon = \frac{E - E_{\rm c}}{W},
  \label{eq_E_rescaled}
\end{align}
where $\hat{I}$ is the unit matrix, and $W$ and $E_{\rm c}$ represent the half-bandwidth and band-center,
\begin{align}
   & W = \frac{E_{\rm max} - E_{\rm min}}{2(1 - \delta)},
  \label{eq_e_width}                                      \\
   & E_{\rm c} = \frac{E_{\rm max} + E_{\rm min}}{2}.
  \label{eq_e_center}
\end{align}
Then, the rescaled Green's functions and Dirac deltas can be expanded as a series of the Chebyshev polynomials, $T_{n} (x)$, as
\begin{align}
   & \delta(\epsilon - \hat{h})
  = \sum_{n = 0}^{\infty} \frac{\Delta_{n}(\epsilon)}{1 + \delta_{n0}} T_{n} (\hat{h}),
  \label{eq_cpdf}
  \\
   & \qquad \Delta_{n}(\epsilon) = \frac{2T_{n}(\epsilon)}{\pi \sqrt{1 - \epsilon^{2}}},
  \label{eq_cpdf_coeff}
\end{align}
and
\begin{align}
   & \hat{g}^{\zeta}(\epsilon, \hat{h})
  =
  \frac{\hbar}{\epsilon+\zeta i\eta-\hat{h}}
  =
  \hbar \sum_{n=0}^{\infty}
  \frac{g_{n}^{\zeta}(\epsilon)}{1+\delta_{n, 0}} T_{n}(\hat{h}),
  \label{eq_cpgf}                                                                                                                                     \\
   & \qquad g_{n}^{\zeta}(\epsilon)=-2 \zeta i \frac{\mathrm{e}^{-\zeta in \arccos (\epsilon + \zeta i\eta)}}{\sqrt{1-(\epsilon + \zeta i\eta)^{2}}},
  \label{eq_cpgf_coeff}
\end{align}
where $\eta \equiv \hbar \gamma / W$ is a dimensionless broadening factor.

By using these polynomial expansions, the thermal average, non-symmetrized linear and second-order responses, $A_{\mu}^{(0)}$, $\tilde{\chi}_{\mu; \alpha}(\omega)$, and $\tilde{\chi}_{\mu; \alpha, \beta}(\omega_{1}, \omega_{2})$ are expressed as
\begin{align}
   & A_{\mu}^{(0)}
  =
  \Lambda^{}_{i}
  \Gamma^{i}_{\mu},
  \label{eq_A_0_ex}                     \\
   & \tilde{\chi}_{\mu; \alpha}(\omega)
  =
  \Lambda_{ij}(\omega;\gamma)
  \Gamma^{ij}_{\mu; \alpha},
  \label{eq_X1_ab_ex}                   \\ &
  \tilde{\chi}_{\mu; \alpha, \beta}(\omega_{1}, \omega_{2})
  =
  \Lambda_{ijk}(\omega_{1}, \omega_{2};\gamma)
  \Gamma^{ijk}_{\mu; \alpha, \beta},
  \label{eq_X2_ab_ex}
\end{align}
where
\begin{widetext}
  \begin{align}
                & \Lambda^{}_{i}
    =
    \sum_{l}
    \frac{c^{(l)}_{i}}{(1 + \delta_{l0})}
    \int_{\mathcal{I}} d \epsilon
    f(\epsilon)
    \Delta^{}_{l}(\epsilon),
    \label{eq_lam_0}                                           \\
                & \Lambda_{ij}(\omega;\gamma)
    =
    \frac{1}{W}
    \sum_{lm}
    \frac{  c^{(l)}_{i}
      c^{(m)}_{j}
    }{(1 + \delta_{l0})(1 + \delta_{m0})}
    \int_{\mathcal{I}} d \epsilon
    f(\epsilon)
    \left[
    g_{l}^{{\rm R}}(\epsilon/\hbar + \tilde{\omega})
    \Delta^{}_{m}(\epsilon)
    +
    \Delta^{}_{l}(\epsilon)
    g_{m}^{{\rm A}}(\epsilon/\hbar - \tilde{\omega})
    \right],
    \label{eq_lam_1}                                           \\
                & \Lambda_{ijk}(\omega_{1}, \omega_{2};\gamma)
    =
    \frac{1}{W^{2}}
    \sum_{lmn}
    \frac{  c^{(l)}_{i}
      c^{(m)}_{j}
      c^{(n)}_{k}
    }{(1+\delta_{l0})(1+\delta_{m0})(1+\delta_{n0})}
    \int_{\mathcal{I}} d \epsilon
    f(\epsilon)
    \left[
    \Delta_{l}(\epsilon)
    g^{{\rm A}}_{m}(\epsilon/\hbar - \tilde{\omega}_{1})
    g^{{\rm A}}_{n}(\epsilon/\hbar - \tilde{\omega}_{1} - \tilde{\omega}_{2})\biggr|_{\eta\to2\eta}
    \right. \cr & \left.
    \hspace{3cm}
    +
    g^{{\rm R}}_{l}(\epsilon/\hbar + \tilde{\omega}_{1})
    \Delta_{m}(\epsilon)
    g^{{\rm A}}_{n}(\epsilon/\hbar - \tilde{\omega}_{2})
    +
    g^{{\rm R}}_{l}(\epsilon/\hbar + \tilde{\omega}_{1} + \tilde{\omega}_{2})\biggr|_{\eta\to2\eta}
    g^{{\rm R}}_{m}(\epsilon/\hbar + \tilde{\omega}_{2})
    \Delta_{n}(\epsilon)
    \right],
    \label{eq_lam_2}
  \end{align}
\end{widetext}
and
\begin{align}
   & \Gamma^{i}_{\mu}
  =
  \mathrm{Tr}
  \left[ \hat{A}_{\mu} \hat{h}^{i} \right],
  \label{eq_gam_0_a}
  \\
   & \Gamma^{ij}_{\mu; \alpha}
  = \mathrm{Tr} \left[ \hat{A}_{\mu} \hat{h}^{i} \hat{B}_{\alpha} \hat{h}^{j} \right],
  \label{eq_gam_1_ab}
  \\
   & \Gamma^{ijk}_{\mu; \alpha, \beta}
  =
  \mathrm{Tr}
  \left[ \hat{A}_{\mu} \hat{h}^{i} \hat{B}_{\alpha} \hat{h}^{j} \hat{B}_{\beta} \hat{h}^{k} \right].
  \label{eq_gam_2_ab}
\end{align}
Here, we have introduced the dimensionless quantities, $\tilde{\omega}_{i} = \omega_{i}/W$, $\tilde{\mu} = \mu / W$, and $\tilde{\beta} = \beta W$, and the rescaled Fermi-Dirac distribution function, $f (\epsilon) = (e^{\tilde{\beta} ( \epsilon - \tilde{\mu})}+1)^{-1}$.
$c_{i}^{(l)}$ is the expansion coefficient in the Chebyshev polynomials, {\it i.e.,} $T_{l}(x)= \sum_{i = 0}^{l} c_{i}^{(l)} x^{i}$.
Once we have obtained the decoupled expressions of the thermal average and response tensors, the formal rescaling procedure in Eqs.~(\ref{eq_H_rescaled}) and (\ref{eq_E_rescaled}) is safely scaled back to the original one.
Namely, we set $W = 1$ and $E_{\rm c} = 0$, hereafter.

\begin{figure}[h]
  \begin{center}
    \includegraphics[width=9cm]{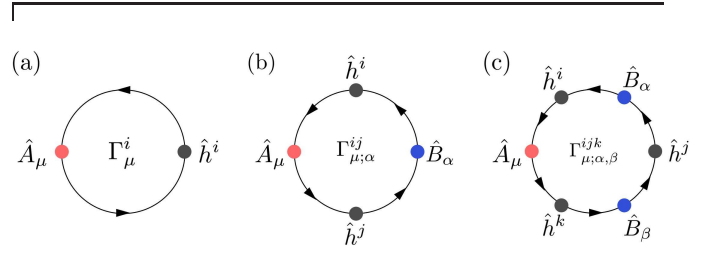}
  \end{center}
  \caption{
    Sequence of the operators in the trace in (a) Eq.~(\ref{eq_gam_0_a}), (b) Eq.~(\ref{eq_gam_1_ab}), and (c) Eq.~(\ref{eq_gam_2_ab}), where the response, external fields, and the power of the Hamiltonian are denoted by the red, blue, and black solid circles, respectively.
  }
  \label{fig_G_diagram}
\end{figure}

The symmetrized expressions of Eqs.~(\ref{eq_A_0_ex})-(\ref{eq_X2_ab_ex}) are the desired results, which are decomposed into two independent parts, $\Lambda$ and $\Gamma$.
The latter part $\Gamma$ depends only on $\hat{A}$, $\hat{B}$, and $\hat{H}_{0}$.
Thus, all the model parameters and symmetry information of the response tensors are embedded in this part $\Gamma$.
The sequence of the operators in $\Gamma$ is depicted in Fig.~\ref{fig_G_diagram}.
By analyzing $\Gamma$, we can identify the essential parameters in the response tensors, which provide a microscopic picture of nontrivial couplings among the electron hopping, SOC, and order parameters, and so on.
On the other hand, the former part $\Lambda$ consists of the external parameters such as frequency, temperature, and chemical potential dependencies through $f (\epsilon)$, and the broadening factor $\gamma$.
In this paper, we mainly focus on the $\Gamma$ part in order to extract the essential parameters.
However, the explicit expression of $\Lambda$ is necessary for those who evaluate the response tensors numerically based on the present formalism.

\subsection{Symmetry of $\Lambda$ and $\Gamma$}
\label{sec:cp_slg}

Here, we discuss the symmetry properties of $\Lambda$ and $\Gamma$, and the expressions of the response tensors are rearranged by using these properties.
The resultant expressions are useful to reduce the relevant terms, especially in the static limit, $\omega=0$.

First, let us consider the symmetry property of $\Lambda$.
In the thermal average $A_{\mu}^{(0)}$, it is obvious that $\Lambda^{}_{i}$ is real.

In the linear response tensors, $\Lambda_{ij}(\omega;\gamma)$ satisfies the following relation,
\begin{align}
   & \Lambda^{*}_{ij}(\omega;\gamma) = \Lambda^{}_{ji} (-\omega;\gamma).
  \label{eq_lam_1_sym}
\end{align}
By introducing $\Lambda_{ij}^{(\pm)} (\omega;\gamma)$ as the even (odd) function of $\omega$:
\begin{align}
  \Lambda_{ij}^{(\pm)} (\omega;\gamma)
      & =
  \frac{\Lambda_{ij}^{} (\omega;\gamma)
    \pm
    \Lambda_{ij}^{} (-\omega;\gamma)}{2}
  \cr &
  =
  \frac{\Lambda_{ij}^{} (\omega;\gamma)
    \pm
    \Lambda_{ji}^{*} (\omega;\gamma)}{2},
  \label{eq_lam1_pm}
\end{align}
we obtain the following relations,
\begin{align}
  \begin{split}
    & \mathrm{Re} \left[
      \Lambda_{ij}^{(\pm)} (\omega;\gamma)
      \right]
    =
    \pm
    \mathrm{Re} \left[
      \Lambda_{ji}^{(\pm)} (\omega;\gamma)
      \right]
    ,
    \\
    & \mathrm{Im} \left[
      \Lambda_{ij}^{(\pm)} (\omega;\gamma)
      \right]
    =
    \mp
    \mathrm{Im} \left[
      \Lambda_{ji}^{(\pm)} (\omega;\gamma)
      \right]
    .
  \end{split}
  \label{eq_lam1_pm_reim}
\end{align}
Namely, $\mathrm{Re}[\Lambda_{ij}^{(+)} (\omega;\gamma)]$ and $\mathrm{Im}[\Lambda_{ij}^{(-)} (\omega;\gamma)]$ are the symmetric tensor, whereas $\mathrm{Im}[\Lambda_{ij}^{(+)} (\omega;\gamma)]$ and $\mathrm{Re}[\Lambda_{ij}^{(-)} (\omega;\gamma)]$ are the anti-symmetric tensor, respectively.

Similarly, $\Lambda^{}_{ijk}(\omega_{1}, \omega_{2};\gamma)$ in the second-order response tensors satisfies the following relation,
\begin{align}
   & \Lambda^{*}_{ijk}(\omega_{1}, \omega_{2};\gamma)
  =
  \Lambda^{}_{kji}(-\omega_{2}, -\omega_{1};\gamma).
  \label{eq_lam_2_sym}
\end{align}
Then, by introducing $\Lambda^{(\pm)}_{ijk}(\omega_{1}, \omega_{2};\gamma)$ as similar to Eq.~(\ref{eq_lam1_pm}) as
\begin{align}
  \Lambda^{(\pm)}_{ijk}(\omega_{1}, \omega_{2};\gamma)
   & =
  \frac{
    \Lambda^{}_{ijk}(\omega_{1}, \omega_{2};\gamma)
    \pm
    \Lambda^{}_{ijk}(-\omega_{1}, -\omega_{2};\gamma)
  }{2}, \cr
   & =
  \frac{
    \Lambda^{}_{ijk}(\omega_{1}, \omega_{2};\gamma)
    \pm
    \Lambda^{*}_{kji}(\omega_{2}, \omega_{1};\gamma)
  }{2},
  \label{eq_lam2_pm}
\end{align}
we obtain the following relations,
\begin{align}
  \begin{split}
    & \mathrm{Re} \left[
      \Lambda^{(\pm)}_{ijk}(\omega_{1}, \omega_{2};\gamma)
      \right]
    =
    \pm
    \mathrm{Re} \left[
      \Lambda^{(\pm)}_{kji}(\omega_{2}, \omega_{1};\gamma)
      \right],
    \\
    & \mathrm{Im} \left[
      \Lambda^{(\pm)}_{ijk}(\omega_{1}, \omega_{2};\gamma)
      \right]
    =
    \mp
    \mathrm{Im} \left[
      \Lambda^{(\pm)}_{kji}(\omega_{2}, \omega_{1};\gamma)
      \right]
    .
  \end{split}
  \label{eq_lam2_pm_reim}
\end{align}
Therefore, $\mathrm{Re}[\Lambda_{ijk}^{(+)} (\omega;\gamma)]$ and $\mathrm{Im}[\Lambda_{ijk}^{(-)} (\omega;\gamma)]$ are symmetric, while $\mathrm{Im}[\Lambda_{ijk}^{(+)} (\omega;\gamma)]$ and $\mathrm{Re}[\Lambda_{ijk}^{(-)} (\omega;\gamma)]$ are anti-symmetric with respect to $(i,\omega_{1})\leftrightarrow (k,\omega_{2})$.

Next, let us discuss the symmetry property of $\Gamma$.
Because $\hat{A}_{\mu}$ and $\hat{B}_{\alpha}$ are the hermitian operators, $\Gamma^{i}_{\mu}$ is real, and $\Gamma^{ij}_{\mu; \alpha}$ and $\Gamma^{ijk}_{\mu; \alpha, \beta}$ satisfy the following relations,
\begin{align}
   & \Gamma^{ij *}_{\mu; \alpha}
  = \Gamma^{ji}_{\mu; \alpha}
  = \Gamma^{ij}_{\alpha; \mu},
  \label{eq_g1_ab_k_sym}
  \\
   & \Gamma^{ijk *}_{\mu; \alpha, \beta}
  = \Gamma^{kji}_{\mu; \beta, \alpha}
  ,
  \label{eq_G2ab_k_sym_1}                \\
   & \Gamma^{ijk}_{\mu; \alpha, \beta}
  = \Gamma^{jki}_{\alpha; \beta, \mu}
  = \Gamma^{kij}_{\beta; \mu, \alpha}
  .
  \label{eq_G2ab_k_sym_2}
\end{align}
Thus, the real (imaginary) part of $\Gamma^{ij}_{\mu; \alpha}$ represents the symmetric (antisymmetric) tensor for $A_{\mu} \leftrightarrow B_{\alpha}$ or $i \leftrightarrow j$:
\begin{align}
   & \mathrm{Re}
  \left[
    \Gamma^{ij}_{\mu; \alpha}
    \right]
  =
  \mathrm{Re}
  \left[
    \Gamma^{ji}_{\mu; \alpha}
    \right]
  =
  \mathrm{Re}
  \left[
    \Gamma^{ij}_{\alpha; \mu}
  \right],       \\
   & \mathrm{Im}
  \left[
    \Gamma^{ij}_{\mu; \alpha}
    \right]
  =
  -\mathrm{Im}
  \left[
    \Gamma^{ji}_{\mu; \alpha}
    \right]
  =
  -\mathrm{Im}
  \left[
    \Gamma^{ij}_{\alpha; \mu}
    \right].
\end{align}

By using Eqs.~(\ref{eq_lam1_pm_reim}), the linear response tensor, Eq.~(\ref{eq_X1_ab_ex}) is rearranged as
\begin{align}
   & \tilde{\chi}_{\mu; \alpha}(\omega)
  =
  \tilde{\chi}_{\mu; \alpha}^{\rm RR}(\omega)
  -
  \tilde{\chi}_{\mu; \alpha}^{\rm II}(\omega)
  \cr
   & \quad\quad\quad\quad
  +
  i \tilde{\chi}_{\mu; \alpha}^{\rm RI}(\omega)
  +
  i \tilde{\chi}_{\mu; \alpha}^{\rm IR}(\omega)
  ,
  \label{eq_chi_1}                                     \\
   & \quad \tilde{\chi}_{\mu; \alpha}^{\rm RR}(\omega)
  =
  \mathrm{Re}
  \left[ \Lambda_{ij}^{(+)} (\omega;\gamma) \right]
  \mathrm{Re}
  \left[ \Gamma^{ij}_{\mu; \alpha} \right],
  \label{eq_X_1_ab_k_RR}
  \\
   & \quad \tilde{\chi}_{\mu; \alpha}^{\rm II}(\omega)
  =
  \mathrm{Im}
  \left[ \Lambda_{ij}^{(+)} (\omega;\gamma) \right]
  \mathrm{Im}
  \left[ \Gamma^{ij}_{\mu; \alpha} \right],
  \label{eq_X_1_ab_k_II}
  \\
   & \quad \tilde{\chi}_{\mu; \alpha}^{\rm RI}(\omega)
  =
  \mathrm{Re}
  \left[ \Lambda_{ij}^{(-)} (\omega;\gamma) \right]
  \mathrm{Im}
  \left[ \Gamma^{ij}_{\mu; \alpha} \right],
  \label{eq_X_1_ab_k_RI}
  \\
   & \quad \tilde{\chi}_{\mu; \alpha}^{\rm IR}(\omega)
  =
  \mathrm{Im}
  \left[ \Lambda_{ij}^{(-)} (\omega;\gamma) \right]
  \mathrm{Re}
  \left[ \Gamma^{ij}_{\mu; \alpha} \right].
  \label{eq_X_1_ab_k_IR}
\end{align}
Note that $\tilde{\chi}_{\mu; \alpha}^{\rm RR}(\omega)$ and $\tilde{\chi}_{\mu; \alpha}^{\rm II}(\omega)$ ($\tilde{\chi}_{\mu; \alpha}^{\rm RI}(\omega)$ and $\tilde{\chi}_{\mu; \alpha}^{\rm IR}(\omega)$) are the even (odd) function of $\omega$.
$\tilde{\chi}_{\mu; \alpha}^{\rm RR}(\omega)$ and $\tilde{\chi}_{\mu; \alpha}^{\rm IR}(\omega)$ are symmetric, whereas the $\tilde{\chi}_{\mu; \alpha}^{\rm II}(\omega)$ and $\tilde{\chi}_{\mu; \alpha}^{\rm RI}(\omega)$ are antisymmetric, for $A_{\mu} \leftrightarrow B_{\alpha}$.

Similarly, by using Eq.~(\ref{eq_lam2_pm_reim}), the second-order response tensor, Eq.~(\ref{eq_X2_ab_ex}) is rearranged as
\begin{align}
                                                                                                                                                                       & \tilde{\chi}_{\mu; \alpha, \beta}(\omega_{1},\omega_{2})
  =
  \tilde{\chi}_{\mu; \alpha, \beta}^{\rm RR}(\omega_{1},\omega_{2})
  -
  \tilde{\chi}_{\mu; \alpha, \beta}^{\rm II}(\omega_{1},\omega_{2})
  \cr
                                                                                                                                                                       & \quad\quad\quad\quad
  +
  i
  \tilde{\chi}_{\mu; \alpha, \beta}^{\rm RI}(\omega_{1},\omega_{2})
  +
  i
  \tilde{\chi}_{\mu; \alpha, \beta}^{\rm IR}(\omega_{1},\omega_{2})
  ,
  \label{eq_chi_2}                                                                                                                                                                                                                                   \\
                                                                                                                                                                       & \,\,\,\, \tilde{\chi}_{\mu; \alpha, \beta}^{\rm RR}(\omega_{1}, \omega_{2})
  =
  \mathrm{Re}
  \left[
    \Lambda_{ijk}^{(+)} (\omega_{1}, \omega_{2};\gamma)
    \right]
  \mathrm{Re}
  \left[
    \Gamma^{ijk}_{\mu; \alpha, \beta}
    \right],
  \label{eq_X_2_ab_k_RR}                                                         \cr                                                                                   &                                                                             \\
                                                                                                                                                                       & \,\,\,\, \tilde{\chi}_{\mu; \alpha, \beta}^{\rm II}(\omega_{1}, \omega_{2})
  =
  \mathrm{Im}
  \left[
    \Lambda_{ijk}^{(+)} (\omega_{1}, \omega_{2};\gamma)
    \right]
  \mathrm{Im}
  \left[
    \Gamma^{ijk}_{\mu; \alpha, \beta}
    \right],
  \label{eq_X_2_ab_k_II}
  \\
                                                                                                                                                                       & \,\,\,\, \tilde{\chi}_{\mu; \alpha, \beta}^{\rm RI}(\omega_{1}, \omega_{2})
  =
  \mathrm{Re}
  \left[
    \Lambda_{ijk}^{(-)} (\omega_{1}, \omega_{2};\gamma)
    \right]
  \mathrm{Im}
  \left[
    \Gamma^{ijk}_{\mu; \alpha, \beta}
    \right],
  \label{eq_X_2_ab_k_RI}                                                                                                                                           \cr &                                                                             \\
                                                                                                                                                                       & \,\,\,\, \tilde{\chi}_{\mu; \alpha, \beta}^{\rm IR}(\omega_{1}, \omega_{2})
  =
  \mathrm{Im}
  \left[
    \Lambda_{ijk}^{(-)} (\omega_{1}, \omega_{2};\gamma)
    \right]
  \mathrm{Re}
  \left[
    \Gamma^{ijk}_{\mu; \alpha, \beta}
    \right].
  \label{eq_X_2_ab_k_IR}
\end{align}
Since $\Lambda^{(-)}_{ij} (\omega;\gamma)$ and $\Lambda^{(-)}_{ijk} (\omega_{1}, \omega_{2};\gamma)$ vanish in the static limit, $\omega, \omega_{1}, \omega_{2} \to 0$, $\tilde{\chi}_{\mu; \alpha}(\omega=0)$ and $\tilde{\chi}_{\mu; \alpha, \beta}(\omega_{1}=0,\omega_{2}=0)$ are real quantities consisting of $\tilde{\chi}^{\rm RR}$ and $\tilde{\chi}^{\rm II}$.

\section{Formula for Systematic Analysis}
\label{sec:params_extraction}

In this section, assuming the translational invariance, we finally provide fundamental momentum-space formula to extract the essential parameters in the thermal average, and linear and nonlinear response tensors.
It is also useful to adopt the real-space formula to grasp nontrivial coupling among electronic degrees of freedom in the response, which is discussed in the supplemental material in detail~\footnote{
  (Supplemental Material) The formula in the real space is discussed, and the properties of a sort of magnetic flux closely related to the real space Berry phase~\cite{Anderson_1955, Ohgushi_2000, Taguchi_2001, Zhang_2020} are clarified.
  Then, the effect of the static magnetic field to the response tensors is also discussed.
}.
In a symmetry-breaking phase without the original translational invariance, the present formula can also be applied by setting the appropriate magnetic unit cell.

\subsection{Hopping Hamiltonian}

Let us consider the hopping Hamiltonian~\cite{Jo_o_2019},
\begin{align}
   & \mathcal{H}_{0} = \sum_{\bm{R}_{e}\bm{R}_{s}}
  [\hat{h}(\bm{r})]^{}_{ab} c_{a}^{\dagger}(\bm{R}_{e}) c_{b}^{}(\bm{R}_{s}),
  \label{eq_H0_real}
\end{align}
where $\bm{R}_{s}$ ($\bm{R}_{e}$) denotes the position vector of the start (end) site of the hopping bond, $\bm{r} = \bm{R}_{e} - \bm{R}_{s}$.
$\hat{h}(\bm{r})$ has the common $\bm{r}$ dependence for all unit cells because of the translational invariance.
The labels, $a$ and $b$, represent the other degrees of freedom of electrons such as the spin, orbital, and sublattice.
Note that $\hat{h}(\bm{r})$ can describe any type of hoppings and potentials including the off-site SOC and mean-field potentials.

Similarly, an hermitian operator $\hat{O}$ is defined by
\begin{align}
  \hat{O}
   & =
  \sum_{\bm{R}_{e}\bm{R}_{s}}
  [\hat{O} (\bm{r})]^{}_{ab} c_{a}^{\dagger}(\bm{R}_{e}) c_{b}^{}(\bm{R}_{s}),
  \label{eq_O_real}
\end{align}
where
\begin{align}
  \hat{O}^{\dagger} (\bm{r}) = \hat{O}^{} (-\bm{r}).
  \label{eq_O_hermite}
\end{align}

In the momentum-space representation, an hermitian operator including the hopping Hamiltonian is given by
\begin{align}
  \hat{O} = \sum_{\bm{k}} [\hat{O} (\bm{k})]^{}_{ab} c^{\dagger}_{a}(\bm{k}) c^{}_{b}(\bm{k}).
  \label{eq_O}
\end{align}
The operator is diagonal in the momentum space because of the translational symmetry, and the momentum- and real-space representations of the operator are related to each other as
\begin{align}
  \hat{O} (\bm{k})
   & =
  \sum_{\bm{r}}  \hat{O} (\bm{r}) e^{-i \bm{k} \cdot \bm{r}}.
  \label{eq_Ok}
\end{align}

\subsection{Momentum-Space Representation}

Let us show how we can identify the essential parameters in the response tensors by using the momentum-space representation of $\Gamma$.
Since all the operators are diagonal in the momentum space, Eqs.~(\ref{eq_gam_0_a})-(\ref{eq_gam_2_ab}) are expressed as
\begin{align}
   & \Gamma^{i}_{\mu} = \sum_{\bm{k}} \Omega^{i}_{\mu} (\bm{k}),
  \label{eq_g0a_1}                                                                                 \\
   & \Gamma^{ij}_{\mu; \alpha} = \sum_{\bm{k}} \Omega^{ij}_{\mu; \alpha} (\bm{k}),
  \label{eq_g1ab_1}                                                                                \\
   & \Gamma^{ijk}_{\mu; \alpha, \beta} = \sum_{\bm{k}} \Omega^{ijk}_{\mu; \alpha, \beta} (\bm{k}),
  \label{eq_g2ab_1}
\end{align}
where we have introduced
\begin{align}
   & \Omega^{i}_{\mu} (\bm{k}) = \mathrm{Tr} \left[\hat{A}_{\mu} (\bm{k}) \hat{h}^{i}(\bm{k})\right],
  \label{eq_g0a_k_1}
  \\
   & \Omega^{ij}_{\mu; \alpha} (\bm{k}) = \mathrm{Tr} \left[\hat{A}_{\mu} (\bm{k}) \hat{h}^{i}(\bm{k}) \hat{B}_{\alpha}(\bm{k}) \hat{h}^{j}(\bm{k})\right],
  \label{eq_g1ab_k_1}
  \\
   & \Omega^{ijk}_{\mu; \alpha, \beta} (\bm{k})
  =
  \mathrm{Tr} \left[\hat{A}_{\mu} (\bm{k}) \hat{h}^{i}(\bm{k}) \hat{B}_{\alpha}(\bm{k}) \hat{h}^{j}(\bm{k}) \hat{B}_{\beta}(\bm{k}) \hat{h}^{k}(\bm{k})\right].
  \label{eq_g2ab_k_1}
\end{align}
Here, $\Omega^{i}_{\mu} (\bm{k})$ is real and characterizes the momentum distribution of the thermal average $A_{\mu}^{(0)}$.
Evaluating $\Omega^{i}_{\mu} (\bm{k})$ analytically, we can identify the effective coupling to $\hat{A}_{\mu}$ at $\bm{k}$ as
\begin{align}
  \hat{H}_{\rm eff}(\bm{k}) \propto
  \Omega^{i}_{\mu} (\bm{k})
  \hat{A}_{\mu} (\bm{k}).
  \label{eq_eff_coupling}
\end{align}

On the other hand, $\Omega^{ij}_{\mu; \alpha} (\bm{k})$ and $\Omega^{ijk}_{\mu; \alpha, \beta} (\bm{k})$ are complex and characterize the momentum distribution of the linear and second-order response tensors.

By evaluating $\Gamma^{i}_{\mu}$, $\Gamma^{ij}_{\mu; \alpha}$, and $\Gamma^{ijk}_{\mu; \alpha, \beta}$, we can identify the essential parameters in the thermal average, linear and second-order response tensors.

Now, let us consider the symmetry property of $\Omega (\bm{k})$.
Suppose that $\hat{A}_{\mu}$, $\hat{B}_{\alpha}$, and $\hat{B}_{\beta}$ have definite parities with respect to the spatial-inversion ($\mathcal{P}$) and time-reversal ($\mathcal{T}$) operations as
\begin{align}
  \begin{split}
    & \mathcal{P}(\hat{A}_{\mu}) = s \hat{A}_{\mu},
    \quad
    \mathcal{T}(\hat{A}_{\mu}) = t \hat{A}_{\mu},
    \quad
    (s, t=\pm 1),                         \\
    & \mathcal{P}(\hat{B}_{\alpha}) = s' \hat{B}_{\alpha},
    \quad
    \mathcal{T}(\hat{B}_{\alpha}) = t' \hat{B}_{\alpha},
    \quad
    (s', t'=\pm 1)
    ,\\
    & \mathcal{P}(\hat{B}_{\beta}) = s'' \hat{B}_{\beta},
    \quad
    \mathcal{T}(\hat{B}_{\beta}) = t'' \hat{B}_{\beta},
    \quad
    (s'', t''=\pm 1).
  \end{split}
\end{align}
Considering the Bloch states, $\hat{h} (\bm{k}) \ket{n \bm{k}} = \epsilon_{n \bm{k}} \ket{n \bm{k}}$, where $n$ denotes the band index, $\ket{n \bm{k}}$ and $\hat{A}_{\mu} (\bm{k})$ are transformed by $\mathcal{P}$ and $\mathcal{T}$ as
\begin{align}
  \begin{split}
    & \mathcal{P} \ket{n \bm{k}} = \ket{n -\bm{k}},
    \quad
    A^{\mu}_{n m} (\bm{k}) = s A^{\mu}_{n m} (-\bm{k}), \\
    & \mathcal{T} \ket{n \bm{k}} = \ket{\bar{n} -\bm{k}},
    \quad
    A^{\mu}_{n m} (\bm{k})=t A^{\mu}_{\bar{m} \bar{n}} (-\bm{k}),
  \end{split}
  \label{eq_A_PT}
\end{align}
where $\bar{n}$ represents the time-reversal partner of $n$.
$\hat{B}_{\alpha} (\bm{k})$ and $\hat{B}_{\beta} (\bm{k})$ are transformed in the same manner.

In the presence of $\mathcal{P}$ ($\epsilon_{n \bm{k}} = \epsilon_{n -\bm{k}}$), $\mathcal{T}$ ($\epsilon_{n \bm{k}} = \epsilon_{\bar{n} -\bm{k}}$), or $\mathcal{P}\mathcal{T}$ ($\epsilon_{n \bm{k}} = \epsilon_{\bar{n} \bm{k}}$) symmetry, we have the following relations:

\underline{$\mathcal{P}$ symmetric}
\begin{align}
   & \Omega^{i}_{\mu} (\bm{k})
  =
  s\, \Omega^{i}_{\mu} (-\bm{k}),
  \label{eq_g0_a_k_P}
  \\&
  \Omega^{ij}_{\mu; \alpha} (\bm{k})
  =
  ss'\, \Omega^{ij}_{\mu; \alpha} (-\bm{k}).
  \label{eq_g1_ab_k_P}
  \\&
  \Omega^{ijk}_{\mu; \alpha, \beta} (\bm{k})
  =
  ss's''\, \Omega^{ijk}_{\mu; \alpha, \beta} (-\bm{k}).
  \label{eq_g2_ab_k_P}
\end{align}

\underline{$\mathcal{T}$ symmetric}
\begin{align}
   & \Omega^{i}_{\mu} (\bm{k})
  =
  t\, \Omega^{i}_{\mu} (-\bm{k}),
  \label{eq_g0_a_k_T}
  \\&
  \Omega^{ij}_{\mu; \alpha} (\bm{k})
  =
  tt'\, \Omega^{ji}_{\mu; \alpha} (-\bm{k})
  =
  tt' \left[\Omega^{ij}_{\mu; \alpha} (-\bm{k})\right]^{*},
  \label{eq_g1_ab_k_T}
  \\&
  \Omega^{ijk}_{\mu; \alpha, \beta} (\bm{k})
  =
  tt't''\, \Omega^{kji}_{\mu; \beta, \alpha} (-\bm{k})
  =
  tt't''\, \left[\Omega^{ijk}_{\mu; \alpha, \beta} (-\bm{k})\right]^{*}.
  \label{eq_g2_ab_k_T}
\end{align}

\underline{$\mathcal{P}\mathcal{T}$ symmetric}
\begin{align}
      & \Omega^{i}_{\mu} (\bm{k})
  =
  st\, \Omega^{i}_{\mu} (\bm{k}),
  \label{eq_g0_a_k_PT}
  \\&
  \Omega^{ij}_{\mu; \alpha} (\bm{k})
  =
  ss'tt'\, \Omega^{ji}_{\mu; \alpha} (\bm{k})
  =
  ss'tt'\, \left[\Omega^{ij}_{\mu; \alpha} (\bm{k})\right]^{*},
  \label{eq_g1_ab_k_PT}
  \\&
  \Omega^{ijk}_{\mu; \alpha, \beta} (\bm{k})
  =
  ss's''tt't''\, \Omega^{kji}_{\mu; \beta, \alpha} (\bm{k})
  \cr & \hspace{2.8cm}
  =
  ss's''tt't''\, \left[\Omega^{ijk}_{\mu; \alpha, \beta} (\bm{k})\right]^{*}.
  \label{eq_g2_ab_k_PT}
\end{align}
By the above symmetry relations, $A_{\mu}^{(0)}$ vanishes when $s = -1$ ($t=-1$) in the $\mathcal{P}$ ($\mathcal{T}$) symmetric system.
In the case of $s s' = -1$ ($s s' s'' = -1$) a broken spatial-inversion symmetry is necessary to obtain a finite $\chi_{\mu; \alpha}$ ($\chi_{\mu; \alpha, \beta}$).
On the other hand, in the case of $t t' = -1$ ($t t' = 1$) a broken time-reversal symmetry is necessary to obtain finite $\chi_{\mu; \alpha}^{\rm RR}$ and $\chi_{\mu; \alpha}^{\rm IR}$ ($\chi_{\mu; \alpha}^{\rm II}$ and $\chi_{\mu; \alpha}^{\rm RI}$), while in the case of $t t' t'' = -1$ ($t t' t'' = 1$) a broken time-reversal symmetry is necessary to obtain finite $\chi_{\mu; \alpha, \beta}^{\rm RR}$ and $\chi_{\mu; \alpha, \beta}^{\rm IR}$ ($\chi_{\mu; \alpha, \beta}^{\rm II}$ and $\chi_{\mu; \alpha, \beta}^{\rm RI}$).

\begin{table*}[t!]
  \begin{center}
    \caption{
      Relevant expressions to extract the essential parameters in the thermal average, and linear and second-order response tensors.
      The operators, $\hat{q}_{\mu}$, $\hat{m}_{\mu}$, $\hat{t}_{\mu}$, and $\hat{g}_{\mu}$ represent the electric odd-parity, magnetic even-parity, magnetic odd-parity, and electric even-parity operators, such as the electric, magnetic, magnetic-toroidal, and electric-toroidal dipole operators.
      $\tau$ is the relaxation time.
      In the columns of ``def.'' and ``KF'', the numbers in the parenthesis indicate the definition of the quantity and the corresponding Kubo formula, respectively.
      It is only necessary to evaluate the real or imaginary part of $\Gamma_{\mu}^{i}$, $\Gamma_{\mu;\alpha}^{ij}$, and $\Gamma_{\mu;\alpha,\beta}^{ijk}$, which is indicated in the column of ``Re/Im''.
      The operators listed in the columns $\Gamma$ represent a set of operators of $\hat{A}_{\mu}$, $\hat{B}_{\alpha}$, $\hat{B}_{\beta}$ in evaluating Eqs.~(\ref{eq_g0a_1})-(\ref{eq_g2ab_k_1}).
    }
    \label{tab_responses}
    \renewcommand{\arraystretch}{1.2}
    \begin{tabular}{c|lc|c|cc|cccc}
      \hline \hline
      $(\mathcal{P}, \mathcal{T}, \mathcal{PT})$    & thermal average / response tensor                       & symbol                                   & $\tau$-dep. & def.                      & KF                      & Re/Im & $\Gamma_{\mu}^{i}$         & $\Gamma_{\mu; \alpha}^{ij}$                                                      & $\Gamma_{\mu; \alpha, \beta}^{ijk}$                  \\ \hline
      \multirow{8}{*}{$(\bigcirc, \times, \times)$} & thermal average                                         & $\braket{\hat{m}_{\mu}}$
                                                    & ---                                                     & (\ref{eq_g0a_1})                         & ---         & Re                        & $\hat{m}_{\mu}$
                                                    & ---                                                     & ---                                                                                                                                                                                                                                                                                         \\
                                                    & linear conductivity, BC term                            & $\sigma_{\mu;\alpha}^{\rm BC}$           & $\tau^{0}$  & (\ref{eq_cond_1_bc_k})    & (\ref{eq_cond_1_bc})    & Im    & ---                        & $(\hat{v}_{\mu}, \hat{v}_{\alpha})$                                              & ---                                                  \\
                                                    & linear E-current induced $\hat{q}_{\mu}$                & $\alpha_{\mu;\alpha}^{({\rm J})}$        & $\tau^{1}$  & (70)                      & (83)                    & Re    & ---                        & $(\hat{q}_{\mu}, \hat{v}_{\alpha})$                                              & ---                                                  \\
                                                    & \hspace{8mm} E-field induced $\hat{t}_{\mu}$            & $\alpha_{\mu;\alpha}^{({\rm E})}$        & $\tau^{0}$  & (71)                      & (84)                    & Im    & ---                        & $(\hat{t}_{\mu}, \hat{v}_{\alpha})$                                              & ---                                                  \\
                                                    & 2nd-order E-current induced $\hat{m}_{\mu}$             & $\alpha_{\mu;\alpha,\beta}^{({\rm JJ})}$ & $\tau^{2}$  & (76)                      & (58)                    & Re    & ---                        & $(\hat{m}_{\mu}, \hat{v}_{\alpha\beta})$                                         & $(\hat{m}_{\mu}, \hat{v}_{\alpha}, \hat{v}_{\beta})$ \\
                                                    & \hspace{13.7mm} E-field induced $\hat{m}_{\mu}$         & $\alpha_{\mu;\alpha,\beta}^{({\rm EE})}$ & $\tau^{0}$  & (78)                      & (66)                    & Re    & ---                        & $(\hat{m}_{\mu}, \hat{v}_{\alpha\beta})$                                         & $(\hat{m}_{\mu}, \hat{v}_{\alpha}, \hat{v}_{\beta})$ \\
                                                    & \hspace{13.7mm} E-current/field induced $\hat{g}_{\mu}$ & $\alpha_{\mu;\alpha,\beta}^{({\rm JE})}$ & $ \tau^{1}$ & (77)                      & (62)                    & Im    & ---                        & $(\hat{g}_{\mu}, \hat{v}_{\alpha\beta})$                                         & $(\hat{g}_{\mu}, \hat{v}_{\alpha}, \hat{v}_{\beta})$ \\
      \hline
      \multirow{7}{*}{$(\times, \bigcirc, \times)$} & thermal average                                         & $\braket{\hat{q}_{\mu}}$
                                                    & ---                                                     & (\ref{eq_g0a_1})                         & ---         & Re                        & $\hat{q}_{\mu}$
                                                    & ---                                                     & ---                                                                                                                                                                                                                                                                                         \\
                                                    & 2nd-order conductivity, BCD term                        & $\sigma_{\mu;\alpha,\beta}^{\rm BCD}$    & $\tau^{1}$  & (\ref{eq_cond_2_bcd_k})   & (\ref{eq_bcd_kubo})     & Im    & ---                        & $(\hat{v}_{\mu}, \hat{v}_{\alpha\beta}), (\hat{v}_{\mu\alpha}, \hat{v}_{\beta})$ & $(\hat{v}_{\mu}, \hat{v}_{\alpha}, \hat{v}_{\beta})$ \\
                                                    & linear E-current induced $\hat{m}_{\mu}$                & $\alpha_{\mu;\alpha}^{({\rm J})}$        & $\tau^{1}$  &                           &                         & Re    & ---                        & $(\hat{m}_{\mu}, \hat{v}_{\alpha})$                                              & ---                                                  \\
                                                    & \hspace{8mm} E-field induced $\hat{g}_{\mu}$            & $\alpha_{\mu;\alpha}^{({\rm E})}$        & $\tau^{0}$  &                           &                         & Im    & ---                        & $(\hat{g}_{\mu}, \hat{v}_{\alpha})$                                              & ---                                                  \\
                                                    & 2nd-order E-current induced $\hat{q}_{\mu}$             & $\alpha_{\mu;\alpha,\beta}^{({\rm JJ})}$ & $\tau^{2}$  &                           &                         & Re    & ---                        & $(\hat{q}_{\mu}, \hat{v}_{\alpha\beta})$                                         & $(\hat{q}_{\mu}, \hat{v}_{\alpha}, \hat{v}_{\beta})$ \\
                                                    & \hspace{13.5mm} E-field induced $\hat{q}_{\mu}$         & $\alpha_{\mu;\alpha,\beta}^{({\rm EE})}$ & $\tau^{0}$  &                           &                         & Re    & ---                        & $(\hat{q}_{\mu}, \hat{v}_{\alpha\beta})$                                         & $(\hat{q}_{\mu}, \hat{v}_{\alpha}, \hat{v}_{\beta})$ \\
                                                    & \hspace{13.5mm} E-current/field induced $\hat{t}_{\mu}$ & $\alpha_{\mu;\alpha,\beta}^{({\rm JE})}$ & $ \tau^{1}$ &                           &                         & Im    & ---                        & $(\hat{t}_{\mu}, \hat{v}_{\alpha\beta})$                                         & $(\hat{t}_{\mu}, \hat{v}_{\alpha}, \hat{v}_{\beta})$ \\
      \hline
      \multirow{8}{*}{$(\times, \times, \bigcirc)$} & thermal average                                         & $\braket{\hat{t}_{\mu}}$
                                                    & ---                                                     & (\ref{eq_g0a_1})                         & ---         & Re                        & $\hat{t}_{\mu}$
                                                    & ---                                                     & ---                                                                                                                                                                                                                                                                                         \\
                                                    & 2nd-order conductivity, Drude term                      & $\sigma_{\mu;\alpha,\beta}^{\rm D}$      & $\tau^{2}$  & (\ref{eq_cond_2_drude_k}) & (\ref{eq_drude_kubo})   & Re    & $\hat{v}_{\mu\alpha\beta}$ & $(\hat{v}_{\mu}, \hat{v}_{\alpha\beta}), (\hat{v}_{\mu\alpha}, \hat{v}_{\beta})$ & $(\hat{v}_{\mu}, \hat{v}_{\alpha}, \hat{v}_{\beta})$ \\
                                                    & \hspace{13.5mm} conductivity, intrinsic term            & $\sigma_{\mu;\alpha,\beta}^{\rm int} $   & $ \tau^{0}$ & (\ref{eq_cond_2_int_k})   & (\ref{eq_int_kubo})     & Re    & ---                        & $(\hat{v}_{\mu}, \hat{v}_{\alpha\beta}), (\hat{v}_{\mu\alpha}, \hat{v}_{\beta})$ & $(\hat{v}_{\mu}, \hat{v}_{\alpha}, \hat{v}_{\beta})$ \\
                                                    & linear E-field induced $\hat{m}_{\mu}$                  & $\alpha_{\mu;\alpha}^{({\rm E})}$        & $\tau^{0}$  &                           &                         & Im    & ---                        & $(\hat{m}_{\mu}, \hat{v}_{\alpha})$                                              & ---                                                  \\
                                                    & \hspace{8mm} E-current induced $\hat{g}_{\mu}$          & $\alpha_{\mu;\alpha}^{({\rm J})}$        & $\tau^{1}$  &                           &                         & Re    & ---                        & $(\hat{g}_{\mu}, \hat{v}_{\alpha})$                                              & ---                                                  \\
                                                    & 2nd-order E-current/field induced $\hat{q}_{\mu}$       & $\alpha_{\mu;\alpha,\beta}^{({\rm JE})}$ & $ \tau^{1}$ &                           &                         & Im    & ---                        & $(\hat{q}_{\mu}, \hat{v}_{\alpha\beta})$                                         & $(\hat{q}_{\mu}, \hat{v}_{\alpha}, \hat{v}_{\beta})$ \\
                                                    & \hspace{13.5mm} E-current induced $\hat{t}_{\mu}$       & $\alpha_{\mu;\alpha,\beta}^{({\rm JJ})}$ & $\tau^{2}$  &                           &                         & Re    & ---                        & $(\hat{t}_{\mu}, \hat{v}_{\alpha\beta})$                                         & $(\hat{t}_{\mu}, \hat{v}_{\alpha}, \hat{v}_{\beta})$ \\
                                                    & \hspace{13.5mm} E-field induced $\hat{t}_{\mu}$         & $\alpha_{\mu;\alpha,\beta}^{({\rm EE})}$ & $\tau^{0}$  &                           &                         & Re    & ---                        & $(\hat{t}_{\mu}, \hat{v}_{\alpha\beta})$                                         & $(\hat{t}_{\mu}, \hat{v}_{\alpha}, \hat{v}_{\beta})$ \\
      \hline
      \multirow{7}{*}{any}                          & thermal average                                         & $\braket{\hat{g}_{\mu}}$
                                                    & ---                                                     & (\ref{eq_g0a_1})                         & ---         & Re                        & $\hat{g}_{\mu}$
                                                    & ---                                                     & ---                                                                                                                                                                                                                                                                                         \\
                                                    & linear conductivity, Drude term                         & $\sigma_{\mu;\alpha}^{\rm D}$            & $\tau^{1}$  & (\ref{eq_cond_1_drude_k}) & (\ref{eq_cond_1_drude}) & Re    & $\hat{v}_{\mu\alpha}$      & $(\hat{v}_{\mu}, \hat{v}_{\alpha})$                                              & ---                                                  \\
                                                    & linear E-field induced $\hat{q}_{\mu}$                  & $\alpha_{\mu;\alpha}^{({\rm E})}$        & $\tau^{0}$  &                           &                         & Im    & ---                        & $(\hat{q}_{\mu}, \hat{v}_{\alpha})$                                              & ---                                                  \\
                                                    & \hspace{8mm} E-current induced $\hat{t}_{\mu}$          & $\alpha_{\mu;\alpha}^{({\rm J})}$        & $\tau^{1}$  &                           &                         & Re    & ---                        & $(\hat{t}_{\mu}, \hat{v}_{\alpha})$                                              & ---                                                  \\
                                                    & 2nd-order E-current/field induced $\hat{m}_{\mu}$       & $\alpha_{\mu;\alpha,\beta}^{({\rm JE})}$ & $ \tau^{1}$ &                           &                         & Im    & ---                        & $(\hat{m}_{\mu}, \hat{v}_{\alpha\beta})$                                         & $(\hat{m}_{\mu}, \hat{v}_{\alpha}, \hat{v}_{\beta})$ \\
                                                    & \hspace{13.5mm} E-current induced $\hat{g}_{\mu}$       & $\alpha_{\mu;\alpha,\beta}^{({\rm JJ})}$ & $\tau^{2}$  &                           &                         & Re    & ---                        & $(\hat{g}_{\mu}, \hat{v}_{\alpha\beta})$                                         & $(\hat{g}_{\mu}, \hat{v}_{\alpha}, \hat{v}_{\beta})$ \\
                                                    & \hspace{13.5mm} E-field induced $\hat{g}_{\mu}$         & $\alpha_{\mu;\alpha,\beta}^{({\rm EE})}$ & $\tau^{0}$  &                           &                         & Re    & ---                        & $(\hat{g}_{\mu}, \hat{v}_{\alpha\beta})$                                         & $(\hat{g}_{\mu}, \hat{v}_{\alpha}, \hat{v}_{\beta})$ \\
      \hline \hline
    \end{tabular}
  \end{center}
\end{table*}

Taking account of the above symmetry properties, the relevant expressions to extract the essential parameters for the thermal average, and linear and second-order response tensors are summarized in Table~\ref{tab_responses}.
The relevant expressions for the linear and second-order conductivities, and the electric-field/current induced response tensors are also shown in the same table, which are explained later in Sect.~\ref{sec:cond}, and in the supplemental material.
For example, in order to extract the essential parameters of the Drude term in the second-order conductivity, we refer to the 7th row in Table~\ref{tab_responses}.
It indicates that it is nonzero only for $(\mathcal{P},\mathcal{T},\mathcal{PT})=(\times,\times,\bigcirc)$ and proportional to $\tau^{2}$ in the clean limit.
Moreover, we need to evaluate the real part of $\Gamma_{\mu}^{i}$ with $\hat{A}_{\mu}=\hat{v}_{\mu\alpha\beta}$, $\Gamma_{\mu;\alpha}^{ij}$ with $(\hat{A}_{\mu},\hat{B}_{\alpha})=(\hat{v}_{\mu},\hat{v}_{\alpha\beta})$, $(\hat{v}_{\mu\alpha},\hat{v}_{\beta})$, and $\Gamma_{\mu;\alpha,\beta}^{ijk}$ with $(\hat{A}_{\mu},\hat{B}_{\alpha},\hat{B}_{\beta})=(\hat{v}_{\mu},\hat{v}_{\alpha},\hat{v}_{\beta})$.
The definitions of $\Gamma_{\mu}^{i}$, $\Gamma_{\mu;\alpha}^{ij}$, and $\Gamma_{\mu;\alpha,\beta}^{ijk}$ are given in Eqs.~(\ref{eq_g0a_1})-(\ref{eq_g2ab_k_1}), and the definitions of the velocity operators, $\hat{v}_{\mu}$, $\hat{v}_{\mu\alpha}$, and $\hat{v}_{\mu\alpha\beta}$ are given in Eqs.~(\ref{eq_velo_k}).

\section{Conductivity}
\label{sec:cond}
When the operators $A$ and $B$ implicitly depend on the external field, they must be further expanded with respect to it, and it gives rise to the additional terms in the response tensors.
As a practical example, we here discuss the linear and second-order electric conductivities in the velocity gauge.
In contrast to the length gauge, the velocity gauge is suitable for our method since the latter expression is explicitly obtained in the analytic form, provided that the explicit analytic form of the Hamiltonian is given.

\subsection{Current Operator in Velocity Gauge}
Let us consider the real-space hopping Hamiltonian given in Eq.~(\ref{eq_H0_real}), and a time-dependent electric field, i.e., $\bm{E}(t) = - \partial_{t} \bm{A}(t)$.
Then, the perturbation Hamiltonian is given by
\begin{align}
   & \mathcal{H}_{\rm ext} (t)
  =
  \sum_{n=1}^{\infty} \frac{e^{n}}{n!} \hat{v}_{\alpha_{1} \ldots \alpha_{n}} A_{\alpha_{1}}(t) \ldots A_{\alpha_{n}}(t),
  \label{eq_pert_vel_gauge}
\end{align}
where $e$ is the elementary charge~\cite{Passos_2018, Parker_2019, Jo_o_2019}, and we have defined the $n$-th order velocity operator:
\begin{align}
   & \hat{v}_{\alpha_{1} \ldots \alpha_{n}}
  =
  \frac{1}{(i \hbar)^{n}} \left[\hat{r}_{\alpha_{n}},\left[\cdots\left[\hat{r}_{\alpha_{1}}, \mathcal{H}_{0} \right]\right] \right],
  \label{eq_velo_1}
\end{align}
where $\hat{r}_{\alpha}$ represents the position operator along $\alpha$ direction.
For $n=1$, $\hat{v}_{\alpha}$ represents the ordinary velocity operator.
The matrix element of the position operator in the real-space basis $\ket{\bm{R} a}$ is given by
\begin{align}
  \braket{\bm{R}_{e} a|\hat{\bm{r}}|\bm{R}_{s} b}
   & =
  \delta_{\bm{R}_{e}, \bm{R}_{s}} \delta_{a b} \bm{R}_{s}
  +
  \bm{d}_{ab}(\bm{R}_{e}, \bm{R}_{s}),
  \label{eq_r_wan}
\end{align}
where $\bm{d}_{ab}(\bm{R}_{e}, \bm{R}_{s})$ is the off-diagonal dipole matrix element~\cite{Pedersen_2001, Foreman_2002, Paul_2003, Sandu_2005, Trani_2005}, which is expected to be small for large distance between $\bm{R}_{s}$ and $\bm{R}_{e}$.
In the present paper, we neglect the second dipole term, since it is hard to evaluate it within the tight-binding model.
The effect of the dipole term and the prescription to effectively include it in the tight-binding model are discussed in the literature~\cite{Pedersen_2001, Foreman_2002, Paul_2003, Sandu_2005, Trani_2005, Wang_wannier_2006, Lee_2018}.
Within this approximation, we use the diagonal and gauge-invariant position operator:
\begin{align}
  \braket{\bm{R}_{e} a|\hat{\bm{r}}|\bm{R}_{s} b}
   & =
  \delta_{\bm{R}_{e}, \bm{R}_{s}} \delta_{a b} \bm{R}_{s}.
  \label{eq_r_diag}
\end{align}

By using Eq.~(\ref{eq_r_diag}), the matrix element of the $n$-th order velocity operator is expressed as
\begin{align}
   & v^{ab}_{\alpha_{1} \ldots \alpha_{n}} (\bm{r})
  =
  \frac{1}{(i \hbar)^{n}} r_{\alpha_{1}} \cdots r_{\alpha_{n}} [\hat{h}(\bm{r})]_{ab}
  \label{eq_velo_real},
\end{align}
where $\bm{r}$ is the bond vector.
Using Eq.~(\ref{eq_Ok}), we obtain the $n$-th order velocity operator in the momentum-space representation as
\begin{align}
   & v^{ab}_{\alpha_{1} \ldots \alpha_{n}} (\bm{k})
  =
  \frac{1}{\hbar^{n}} \partial_{\alpha_{1}} \cdots \partial_{\alpha_{n}} [\hat{h} (\bm{k})]_{ab}
  \label{eq_velo_k},
\end{align}
where $\partial_{\alpha}$ stands for the $k_{\alpha}$ derivative.
Both Eqs.~(\ref{eq_velo_real}) and (\ref{eq_velo_k}) can be obtained analytically, provided an analytic form of the hopping Hamiltonian.

The current operator in the velocity gauge is defined by the commutator of the position operator and the total Hamiltonian as
\begin{align}
      & \hat{j}_{\mu} (t) =
  -\frac{e}{V}\frac{1}{i\hbar} \left[\hat{r}_{\mu}, \mathcal{H} (t)\right],
  \cr
      & =
  -\frac{e}{V}
  \left(
  \hat{v}_{\mu} + e \hat{v}_{\mu\alpha} A_{\alpha}(t)
  +
  \frac{e^{2}}{2!} \hat{v}_{\mu\alpha\beta} A_{\alpha}(t) A_{\beta}(t)
  +\cdots\right),
  \cr &
  \label{eq_j_t}
\end{align}
where $V$ is the system volume~\cite{Parker_2019}.
By using the relation $\bm{E}(\omega) = i \omega \bm{A}(\omega)$, the current operator in the frequency domain is obtained as
\begin{align}
   & \hat{j}_{\mu}(\omega)
  =
  \sum_{n = 0} \hat{j}_{\mu}(\omega; n)
  ,
  \label{eq_j_om}
  \\&\qquad
  \hat{j}_{\mu}(\omega; n)
  =
  -\frac{2\pi}{V} \frac{e^{n+1}}{n !} \hat{v}_{\mu\alpha_{1} \ldots \alpha_{n}}
  \cr
   & \qquad\qquad \times
  \left(
  \prod_{i=1}^{n} \int_{\omega_{i}} \frac{E_{\alpha_{i}}(\omega_{i})}{i\omega_{i}}
  \right)
  \delta \left(\omega - \omega_{[n]}\right).
  \label{eq_j_om_n}
\end{align}
The explicit forms of $\hat{j}_{\mu}(\omega; n)$ up to $n = 2$ are given by
\begin{align}
  \hat{j}_{\mu}(\omega; 0)
   & =
  -\frac{2\pi}{V}e\hat{v}_{\mu}\delta(\omega),
  \label{eq_j_om_0}
  \\
  \hat{j}_{\mu}(\omega; 1)
   & =
  -\frac{e^{2}}{V} \hat{v}_{\mu\alpha} \frac{E_{\alpha}(\omega)}{i\omega},
  \label{eq_j_om_1}
  \\
  \hat{j}_{\mu}(\omega; 2)
   & =
  -\frac{\pi e^{3}}{V}
  \hat{v}_{\mu\alpha\beta}
  \int_{\omega_{1},\omega_{2}}
  \frac{E_{\alpha}(\omega_{1})}{i\omega_{1}} \frac{E_{\beta}(\omega_{2})}{i\omega_{2}}
  \cr
   & \qquad\qquad\qquad \times
  \delta \left(\omega - \omega_{1} - \omega_{2}\right).
  \label{eq_j_om_2}
\end{align}

Equation~(\ref{eq_pert_vel_gauge}) in the frequency domain is given by
\begin{align}
  \mathcal{H}_{\rm ext}(\omega)
      & = 2\pi \sum_{n=1}^{\infty} \frac{e^{n}}{n !} \hat{v}_{\alpha_{1} \ldots \alpha_{n}}
  \cr & \,\,\, \times
  \left(\prod_{i=1}^{n} \int_{\omega_{i}} \frac{E_{\alpha_{i}}(\omega_{i})}{i\omega_{i}}\right)
  \delta (\omega - \omega_{[n]}).
  \label{eq_Vom}
\end{align}
By comparing Eq.~(\ref{eq_Vom}) with
\begin{align}
  \mathcal{H}_{\rm ext}(\omega)
   & =
  \int_{\omega'} \hat{B}_{\alpha} (\omega - \omega') E_{\alpha}  (\omega'),
  \label{eq_HextBomE}
\end{align}
the corresponding operator $\hat{B}$ in Eq.~(\ref{eq_HextBomE}) is obtained as
\begin{align}
   & \hat{B}_{\alpha}(\omega - \omega')
  =
  \sum_{n=0} \hat{B}_{\alpha}(\omega -  \omega'; n),
  \label{eq_B_om}
  \\
   & \hat{B}_{\alpha}(\omega -  \omega'; n)
  =
  -V \frac{1}{i\omega'} \frac{1}{n+1} \hat{j}_{\alpha}(\omega - \omega'; n).
  \label{eq_B_om_n}
\end{align}
As shown in the above, the current operator itself involves the external electric fields, and hence the additional expansion 
is needed to obtain the response tensors, which gives the additional terms to Eqs.~(\ref{eq_X1_ab}) and (\ref{eq_X2_ab}).
Then, the linear and second-order conductivities are defined by
\begin{multline}
  j_{\mu} (\omega) =
  \sigma_{\mu; \alpha}(\omega) E_{\alpha} (\omega) \cr
  + \int_{\omega_{1}, \omega_{2}} \sigma_{\mu; \alpha, \beta}(\omega_{1}, \omega_{2}) E_{\alpha} (\omega_{1}) E_{\beta} (\omega_{2}) \cr
  \times 2\pi \delta(\omega - \omega_{1} - \omega_{2}) + \cdots,
\end{multline}
and the explicit expressions of the conductivity tensors are given below.
A detailed derivation of the nonlinear conductivity is shown in Appendix~\ref{sec:nlc}.

\subsection{Linear Conductivity}
\label{sec:cond_1}
In the clean limit, $\tau = \gamma^{-1} \to \infty$, the linear DC conductivity obtained from the standard Kubo formula is decomposed into the Drude and BC terms as
\begin{align}
  \sigma^{}_{\mu;\alpha}
   & =
  \sigma^{\rm D}_{\mu;\alpha} + \sigma^{\rm BC}_{\mu;\alpha}.
  \label{eq_cond_1}
\end{align}
These two components in the velocity gauge are explicitly given by (see Appendix~\ref{sec:lcon_vel})
\begin{align}
  \sigma^{\rm D}_{\mu;\alpha}
   & =
  \frac{e^2}{\gamma \hbar^{2}V} \sum_{\bm{k}n}
  f_{n \bm{k}} \partial_{\mu} \partial_{\alpha} \epsilon_{n\bm{k}},
  \label{eq_cond_1_drude} \\
  \sigma^{\rm BC}_{\mu; \alpha}
   & =
  -\frac{e^{2}}{\hbar V} \sum_{\bm{k}n}
  \epsilon_{\mu\alpha\beta} f_{n \bm{k}} \Omega^{\beta}_{n} (\bm{k}),
  \label{eq_cond_1_bc}
\end{align}
where $f_{n\bm{k}}=f(\epsilon_{n\bm{k}})$, and $\Omega^{\kappa}_{n} (\bm{k})$ is the Berry curvature in the momentum space:
\begin{align}
   & \Omega^{\kappa}_{n} (\bm{k})
  =
  i \sum_{m \neq n}
  \epsilon_{\kappa \alpha \beta} \xi^{nm}_{\alpha} (\bm{k}) \xi^{mn}_{\beta} (\bm{k}).
  \label{eq_bc}
\end{align}
Here, $\xi^{nm}_{\alpha} (\bm{k}) = i \braket{n \bm{k}| \partial_{\alpha} |m \bm{k}}$ is the Berry connection.
$\sigma^{\rm D}_{\mu;\alpha}$ ($\sigma^{\rm BC}_{\mu;\alpha}$) is the dissipative (non-dissipative) part in the clean limit.
In the linear conductivity tensor, the symmetric part is the Drude term, i.e., $\sigma^{\rm D}_{\mu;\alpha} = \sigma^{\rm D}_{\alpha;\mu}$, while the antisymmetric part is the BC term, $\sigma^{\rm BC}_{\mu;\alpha} = - \sigma^{\rm BC}_{\alpha;\mu}$.
The relation among the conditions of $\mathcal{P}$, $\mathcal{T}$, and $\mathcal{PT}$ symmetries for nonzero $\sigma^{\rm D}_{\mu;\alpha}$ and $\sigma^{\rm BC}_{\mu;\alpha}$ is summarized in Table~\ref{tab_responses}.
Note that $\sigma^{\rm D}_{\mu;\alpha}$ always exists irrespective of these symmetries, while the broken $\mathcal{T}$ symmetry is necessary to obtain finite $\sigma^{\rm BC}_{\mu;\alpha}$.

On the other hand, the Chebyshev polynomial expansion of the linear conductivity is given as follows.
By substituting Eqs.~(\ref{eq_j_om}) and (\ref{eq_B_om}) into Eq.~(\ref{eq_Ao_n_ex}), the linear conductivity is obtained as (see Appendix~\ref{sec:nlc} in detail)
\begin{align}
   & \sigma_{\mu; \alpha}(\omega)
  =
  \frac{ie^{2}}{V (\omega + i\gamma)}
  \left\{
  v_{\mu\alpha}^{(0)} + \chi_{\mu; \alpha} (\omega)
  \right\}.
  \label{eq_cond_1_om}
\end{align}
Taking the static limit, $\omega \to 0$, the symmetric tensor $\sigma^{\rm D}_{\mu;\alpha}$ and the antisymmetric tensor $\sigma^{\rm BC}_{\mu;\alpha}$ are obtained as
\begin{align}
   & \sigma^{\rm D}_{\mu;\alpha}
  =
  \frac{e^{2}}{\gamma V}
  \left\{
  v_{\mu\alpha}^{(0)} + \chi_{\mu; \alpha}^{\rm RR} (0)
  \right\}
  ,
  \label{eq_cond_1_drude_ex}      \\
   & \sigma^{\rm BC}_{\mu;\alpha}
  =
  - \frac{e^{2}}{V}
  \lim_{\gamma\to+0}
  \frac{d}{d \gamma} \chi_{\mu; \alpha}^{\rm II}(\gamma)
  ,
  \label{eq_cond_1_bc_ex}
\end{align}
where $v_{\mu\alpha}^{(0)}$, $\chi_{\mu; \alpha}^{\rm RR}$, and $\chi_{\mu; \alpha}^{\rm II}$ are given by Eqs.~(\ref{eq_A_0_ex}), (\ref{eq_X_1_ab_k_RR}), and (\ref{eq_X_1_ab_k_II}).
We have used the fact that the Drude and BC terms are proportional to $\gamma^{-1}$ and $\gamma^{0}$, respectively.

Using Eqs.~(\ref{eq_g0a_1}) and (\ref{eq_g1ab_1}), Eqs.~(\ref{eq_cond_1_drude_ex}) and (\ref{eq_cond_1_bc_ex}) are expressed as
\begin{align}
   & \sigma^{\rm D}_{\mu;\alpha}
  =
  \frac{e^{2}}{\gamma V}
  \left\{
  \Lambda^{}_{i} \Gamma^{i}_{\mu\alpha}
  +
  \mathrm{Re} \left[ \Lambda_{ij}^{(+,0)} \right]
  \mathrm{Re} \left[ \Gamma^{ij}_{\mu;\alpha} \right]
  \right\},
  \label{eq_cond_1_drude_k}       \\
   & \sigma^{\rm BC}_{\mu;\alpha}
  =
  -\frac{e^{2}}{V}
  \mathrm{Im}
  \left[ \Lambda_{ij}^{(+,1)} \right]
  \mathrm{Im}
  \left[ \Gamma^{ij}_{\mu;\alpha} \right]
  ,
  \label{eq_cond_1_bc_k}
\end{align}
where
\begin{align}
   & \Gamma^{i}_{\mu\alpha} = \sum_{\bm{k}} \mathrm{Tr} \left[\hat{v}_{\mu\alpha}(\bm{k}) \hat{h}^{i}(\bm{k})\right],
  \label{eq_g0vab}                                                                                                                                             \\
   & \Gamma^{ij}_{\mu;\alpha} = \sum_{\bm{k}} \mathrm{Tr} \left[\hat{v}_{\mu}(\bm{k}) \hat{h}^{i}(\bm{k}) \hat{v}_{\alpha}(\bm{k}) \hat{h}^{j}(\bm{k})\right].
  \label{eq_g1vab}
\end{align}
Here we have introduced the derivative of $\Lambda_{ij}^{(+)}(0;\gamma)$ as
\begin{align}
   & \Lambda_{ij}^{(+,n)}
  =
  \lim_{\gamma\to+0}
  \frac{1}{n!} \frac{d^{n}}{d \gamma^{n}} \Lambda_{ij}^{(+)}(0;\gamma).
  \label{eq_lamd_1n}
\end{align}

\subsection{Second-Order Conductivity}
\label{sec:cond_2}
Similarly, the second-order DC conductivity is known to be classified by three terms as
\begin{align}
   & \sigma_{\mu; \alpha, \beta}
  =
  \sigma_{\mu; \alpha, \beta}^{\rm D}
  +\sigma_{\mu; \alpha, \beta}^{\rm BCD}
  +\sigma_{\mu; \alpha, \beta}^{\rm int},
\end{align}
which represent the Drude, BCD, and intrinsic terms, respectively~\cite{HW_nonlinear_cond_2020}.
Each term is explicitly given by~\footnote{(Supplemental Material) Derivation of the explicit form of the second-order conductivity in the velocity gauge is given.}
\begin{align}
   & \sigma_{\mu; \alpha, \beta}^{\rm D}
  =
  -\frac{e^{3}}{2\gamma^{2}\hbar^{3} V}
  \sum_{\bm{k}n}
  f_{n\bm{k}} \partial_{\mu}\partial_{\alpha}\partial_{\beta} \epsilon_{n \bm{k}},
  \label{eq_drude_kubo}                    \\
   & \sigma_{\mu; \alpha, \beta}^{\rm BCD}
  =
  \frac{e^{3}}{2\gamma\hbar^{2} V}
  \sum_{\bm{k}n}
  f_{n\bm{k}} \epsilon_{\mu \alpha \kappa} D_{n}^{\beta \kappa}(\bm{k})
  +
  \left[\alpha \leftrightarrow \beta\right],
  \label{eq_bcd_kubo}                      \\
   & \sigma_{\mu; \alpha, \beta}^{\rm int}
  =
  \frac{e^{3}}{\hbar V} \sum_{\bm{k}n \neq m} \left[
  \frac{1}{2}\frac{f_{n \bm{k}} - f_{m \bm{k}}}{(\epsilon_{n\bm{k}} - \epsilon_{\bm{k}m})^{2}}g_{\alpha\beta}^{nm}(\bm{k})
  \partial_{\mu} \epsilon_{n\bm{k}}
  \right.  \cr
   & \quad \left.
  +2 \left(-\frac{\partial f_{n \bm{k}}}{\partial \epsilon_{n\bm{k}}}\right)
  \partial_{\alpha} \epsilon_{n\bm{k}}
  \frac{g_{\mu \beta}^{nm}(\bm{k})}{(\epsilon_{n\bm{k}} - \epsilon_{\bm{k}m})}
  \right] + \left[\alpha \leftrightarrow \beta\right],
  \label{eq_int_kubo}
\end{align}
where the Berry curvature dipole $D_{n}^{\alpha \beta} (\bm{k})$~\cite{Sodemann_bcd_2015} is given by
\begin{align}
   & D_{n}^{\alpha \beta} (\bm{k})
  =
  \partial_{\alpha}
  \Omega_{n}^{\beta}(\bm{k}),
  \label{eq_bcd}
\end{align}
and $g_{\alpha\beta}^{nm}(\bm{k})$ is the quantum metric~\cite{Gao_qm_2020, HW_nonlinear_optoele_2021}:
\begin{align}
  g_{\alpha\beta}^{nm}(\bm{k})
   & =
  \frac{1}{2} \left(
  \xi_{\alpha}^{nm}(\bm{k}) \xi_{\beta}^{mn}(\bm{k}) + \xi_{\beta}^{nm}(\bm{k}) \xi_{\alpha}^{mn}(\bm{k})
  \right).
  \label{eq_qm}
\end{align}
In the clean limit, both $\sigma^{\rm D}_{\mu;\alpha, \beta}$ and $\sigma^{\rm BCD}_{\mu;\alpha, \beta}$ are the dissipative parts, whereas $\sigma^{\rm int}_{\mu;\alpha, \beta}$ is insensitive of $\gamma$, and hence, this term is negligible in good metals.

The conditions of $\mathcal{P}$, $\mathcal{T}$, and $\mathcal{PT}$ symmetries for nonzero $\sigma^{\rm D}_{\mu;\alpha,\beta}$, $\sigma^{\rm BCD}_{\mu;\alpha,\beta}$, and $\sigma^{\rm int}_{\mu;\alpha,\beta}$ are also summarized in Table~\ref{tab_responses}~\cite{HW_nonlinear_cond_2020}.
The broken $\mathcal{P}$ symmetry is necessary for all terms to be finite.
In the $\mathcal{P}$-broken system, $\sigma^{\rm D}_{\mu;\alpha, \beta}$ and $\sigma^{\rm int}_{\mu;\alpha, \beta}$ are allowed to be finite in the $\mathcal{PT}$ symmetric system, while $\sigma^{\rm BCD}_{\mu;\alpha}$ is allowed to be finite in the $\mathcal{T}$ symmetric system.

Similar to Eq.~(\ref{eq_cond_1_om}), the Chebyshev polynomial expansion of the second-order conductivity is given by (see Appendix~\ref{sec:nlc} in detail)
\begin{widetext}
  \begin{multline}
    \sigma_{\mu; \alpha, \beta}(\omega_{1}, \omega_{2})
    =
    \frac{e^{3}}{2V (\omega_{1} + i \gamma) (\omega_{2} + i \gamma)}
    \biggl\{
    \frac{1}{2} v_{\mu\alpha\beta}^{(0)}
    +
    \frac{1}{2} \tilde{\chi}_{\mu; (\alpha\beta)}(\omega_{1} + \omega_{2})
    +
    \tilde{\chi}_{(\mu\alpha); \beta}(\omega_{2})
    +
    \tilde{\chi}_{\mu; \alpha, \beta}(\omega_{1}, \omega_{2})
    \biggr\}
    \\
    + [(\alpha, \omega_{1}) \leftrightarrow (\beta, \omega_{2})].
    \label{eq_cond_2}
  \end{multline}
  Note that $\tilde{\chi}_{\mu; (\alpha\beta)} = \chi_{\mu; (\alpha\beta)}$ and $\tilde{\chi}_{(\mu\alpha); \beta} = \chi_{(\mu\alpha); \beta}$.
  We consider the case of $\mathcal{PT}$-symmetric and $\mathcal{T}$-symmetric systems separately as follows.

  \subsubsection{$\mathcal{PT}$ symmetric system}
  In this case, the BCD term vanishes.
  Taking the static limit, $\omega_{1}, \omega_{2} \to 0$, the Drude and intrinsic terms are given by
  \begin{align}
     & \sigma^{\rm D}_{\mu;\alpha, \beta}
    =
    - \frac{e^{3}}{2\gamma^{2}V} \lim_{\gamma\to+0}
    \left\{
    \frac{1}{2} v_{\mu\alpha\beta}^{(0)}
    +
    \frac{1}{2} \tilde{\chi}_{\mu; (\alpha\beta)}^{\rm RR}(2 \gamma)
    +
    \tilde{\chi}_{(\mu\alpha); \beta}^{\rm RR}(\gamma)
    +
    \tilde{\chi}_{\mu; \alpha,\beta}^{\rm RR}(\gamma, \gamma)
    \right\}
    + [\alpha \leftrightarrow \beta]
    ,
    \label{eq_cond_2_drude_ex}              \\
     & \sigma^{\rm int}_{\mu;\alpha, \beta}
    =
    - \frac{e^{3}}{4 V} \lim_{\gamma\to+0} \frac{d^{2}}{d \gamma^{2}}
    \left\{
    \frac{1}{2} \tilde{\chi}_{\mu; (\alpha\beta)}^{\rm RR}(2 \gamma)
    +
    \tilde{\chi}_{(\mu\alpha); \beta}^{\rm RR}(\gamma)
    +
    \tilde{\chi}_{\mu; \alpha,\beta}^{\rm RR}(\gamma, \gamma)
    \right\}
    + [\alpha \leftrightarrow \beta]
    ,
    \label{eq_cond_2_int_ex}
  \end{align}
  where we have used the fact that the Drude and intrinsic terms are proportional to $\gamma^{-2}$ and $\gamma^{0}$, respectively.

  The explicit expressions of Eqs.~(\ref{eq_cond_2_drude_ex}) and (\ref{eq_cond_2_int_ex}) in terms of $\Lambda$ and $\Gamma$ are given by
  \begin{align}
     & \sigma^{\rm D}_{\mu;\alpha, \beta}
    =
    -
    \frac{e^{3}}{4\gamma^{2} V}
    \left\{
    \Lambda^{}_{i} \Gamma^{i}_{\mu\alpha\beta}
    +
    \mathrm{Re} \left[ \Lambda_{ij}^{(+,0)} \right]
    \left(
    \mathrm{Re} \left[ \Gamma^{ij}_{\mu; (\alpha\beta)} \right]
    +
    2\,\mathrm{Re}
    \left[ \Gamma^{ij}_{(\mu\alpha); \beta}\right]
    \right)
    +
    2\,\mathrm{Re}
    \left[ \Lambda_{ijk}^{(+,0)} \right]
    \mathrm{Re}
    \left[ \Gamma^{ijk}_{\mu; \alpha, \beta} \right]
    \right\}
    + [\alpha \leftrightarrow \beta],
    \label{eq_cond_2_drude_k}               \\
     & \sigma^{\rm int}_{\mu;\alpha, \beta}
    =
    -\frac{e^{3}}{4V}
    \left\{
    \mathrm{Re} \left[ \Lambda_{ij}^{(+,2)} \right]
    \left(
    2\, \mathrm{Re} \left[ \Gamma^{ij}_{\mu; (\alpha\beta)} \right]
    +
    \mathrm{Re} \left[ \Gamma^{ij}_{(\mu\alpha); \beta}\right]
    \right)
    +
    \mathrm{Re}
    \left[ \Lambda_{ijk}^{(+,2)} \right]
    \mathrm{Re} \left[\Gamma^{ijk}_{\mu; \alpha, \beta} \right]
    \right\}
    + [\alpha \leftrightarrow \beta],
    \label{eq_cond_2_int_k}
  \end{align}
  where
  \begin{align}
     & \Gamma^{i}_{\mu\alpha\beta} = \sum_{\bm{k}} \mathrm{Tr} \left[\hat{v}_{\mu\alpha\beta}(\bm{k}) \hat{h}^{i}(\bm{k})\right],
    \label{eq_g0vabc}                                                                                                                                                                                                 \\
     & \Gamma^{ij}_{\mu; (\alpha\beta)} = \sum_{\bm{k}} \mathrm{Tr} \left[\hat{v}_{\mu}(\bm{k}) \hat{h}^{i}(\bm{k}) \hat{v}_{\alpha\beta}(\bm{k}) \hat{h}^{j}(\bm{k})\right],
    \label{eq_g1va_bc}                                                                                                                                                                                                \\
     & \Gamma^{ij}_{(\mu\alpha); \beta} = \sum_{\bm{k}} \mathrm{Tr} \left[\hat{v}_{\mu\alpha}(\bm{k}) \hat{h}^{i}(\bm{k}) \hat{v}_{\beta}(\bm{k}) \hat{h}^{j}(\bm{k})\right],
    \label{eq_g1vab_c}                                                                                                                                                                                                \\
     & \Gamma^{ijk}_{\mu; \alpha, \beta} = \sum_{\bm{k}} \mathrm{Tr} \left[\hat{v}_{\mu}(\bm{k}) \hat{h}^{i}(\bm{k}) \hat{v}_{\alpha}(\bm{k}) \hat{h}^{j}(\bm{k}) \hat{v}_{\beta}(\bm{k}) \hat{h}^{k}(\bm{k})\right],
    \label{eq_g2vabc}
  \end{align}
  and
  \begin{align}
     & \Lambda_{ijk}^{(+,n)}
    =
    \lim_{\gamma\to+0}
    \frac{1}{n!}
    \frac{d^{n}}{d \gamma^{n}}
    \Lambda_{ijk}^{(+)}(0, 0;\gamma)
    .
    \label{eq_lamd_2n}
  \end{align}

  \subsubsection{$\mathcal{T}$ symmetric system}

  In this case, the Drude and intrinsic terms vanish.
  The BCD term in the static limit is given by
  \begin{align}
    \sigma^{\rm BCD}_{\mu;\alpha, \beta}
    =
    \frac{e^{3}}{2\gamma V}
    \lim_{\gamma\to+0}
    \frac{d}{d \gamma}
    \left\{
    \frac{1}{2}
    \tilde{\chi}_{\mu; (\alpha\beta)}^{\rm II}(2 \gamma)
    +
    \tilde{\chi}_{(\mu\alpha); \beta}^{\rm II}(\gamma)
    +
    \tilde{\chi}_{\mu; \alpha, \beta}^{\rm II}(\gamma, \gamma)
    \right\}
    + [\alpha \leftrightarrow \beta]
    ,
    \label{eq_cond_2_bcd_ex}
  \end{align}
  where we have used the fact that the BCD term is proportional to $\gamma^{-1}$.

  The BCD term in in terms of $\Lambda$ and $\Gamma$ is given by
  \begin{align}
    \sigma^{\rm BCD}_{\mu;\alpha, \beta}
    =
    \frac{e^{3}}{2\gamma V}
    \left\{
    \mathrm{Im}
    \left[
      \Lambda_{ij}^{(+,1)}
      \right]
    \left(
    \mathrm{Im}
    \left[
      \Gamma^{ij}_{\mu; (\alpha\beta)}
      \right]
    +
    \mathrm{Im}
    \left[
      \Gamma^{ij}_{(\mu\alpha); \beta}
      \right]
    \right)
    +
    \mathrm{Im}
    \left[
      \Lambda_{ijk}^{(+,1)}
      \right]
    \mathrm{Im}
    \left[
      \Gamma^{ijk}_{\mu; \alpha, \beta}
      \right]
    \right\}
    + [\alpha \leftrightarrow \beta]
    .
    \label{eq_cond_2_bcd_k}
  \end{align}
\end{widetext}

\section{Application to Ferroelectric Tin Telluride Monolayer}
\label{sec:snte}

We demonstrate the present analysis method by taking a specific example, the ferroelectric SnTe monolayer system.
First, we construct the hopping Hamiltonian by means of the symmetry-adopted multipole basis in \ref{sec:snte_tb}.
This procedure is useful to construct automatically the hopping Hamiltonian and analyze response tensors by combining it with computational analysis.
Then, we discuss the anti-symmetric spin and orbital splittings in \ref{sec:antisplit}, and nonlinear Hall effect in \ref{sec:nlhesnte}.

\subsection{Two-Orbital Hopping Model}
\label{sec:snte_tb}

\begin{figure}[t]
  \begin{center}
    \includegraphics[width=8.5cm]{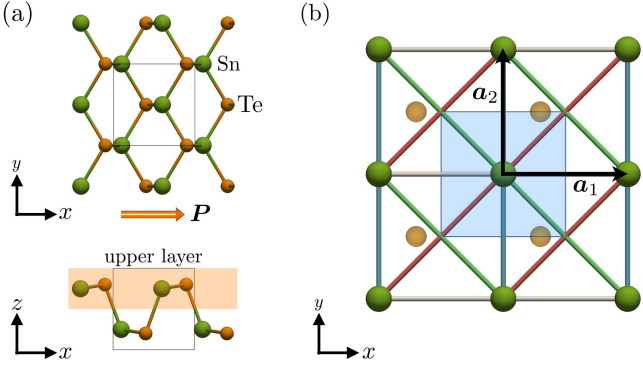}
  \end{center}
  \caption{
    Crystal structure of the ferroelectric monolayer SnTe ($x$ principal axis),
    (a) The top ($z$-axis) and side ($y$-axis) views.
    The in-plane atomic displacement of Sn (green sphere) and Te (orange sphere) gives rise to the electric polarization $\bm{P}$ (orange arrow).
    The orange shaded area represents the upper layer.
    (b) The simplified effective model ($y$ principal axis) by taking the upper layer of SnTe.
    $\bm{a}_{1}$ and $\bm{a}_{2}$ are the unit vectors.
    The square represents the unit cell.
  }
  \label{fig_snte_crystal}
\end{figure}

The monolayer SnTe exhibits the in-plane spontaneous electric polarization below the transition temperature $T_{\rm c} = 270$ K~\cite{snte_Chang_2016}.
The {\it ab initio} calculation predicts that it exhibits the nonlinear Hall effect~\cite{snte_kim_2019}.
The space group of the monolayer SnTe in the ferroelectric phase is $Pmn2_{1}$ (\#31, $C_{2v}^{7}$) with the mirror symmetry in the $xz$ plane, and glide symmetry with $xy$ plane as shown in Fig.~\ref{fig_snte_crystal}(a)~\cite{snte_crystal_fp_2017}.

The opposite in-plane displacement of Sn and Te atoms along the $x$-axis gives rise to the ferroelectric polarization $\bm{P}$.
Near the conduction (valence) band edge mainly consists of Sn (Te) $p_{x}$ and $p_{y}$ orbitals~\cite{snte_kim_2019}.
By focusing on the conduction band edge, we consider the simplified two-dimensional hopping model with Sn (green sphere) $p_{x}$ and $p_{y}$ orbitals by taking the upper layer of the SnTe as shown in Fig.~\ref{fig_snte_crystal}(b).
The presence of Te atoms (orange sphere) affects the effective hopping parameters between Sn atoms.
The space group of the effective model is $P_{m}$ (\#6, $C_{s}^{1}$), and the corresponding point-group symmetry is $C_{s}$.

The hopping Hamiltonian is generally expressed as
\begin{align}
  \mathcal{H}_{0}
  =
  \sum_{\bm{r}, \bm{R}_{s}}
  [\hat{h} (\bm{r})]^{\sigma\sigma'}_{ab}
  c_{a \sigma }^{\dagger}(\bm{r} + \bm{R}_{s}) c_{b  \sigma'}^{}(\bm{R}_{s}),
  \label{eq_H0_r}
\end{align}
where $c_{a\sigma}^{\dagger}(\bm{r} + \bm{R}_{s})$ ($c_{a\sigma}^{}(\bm{R}_{s})$) is the creation (annihilation) operator of the position $\bm{r} + \bm{R}_{s}$ ($\bm{R}_{s}$), orbital $a = p_{x}, p_{y}$, and spin $\sigma = \uparrow, \downarrow$.
Following the bottom-up procedure for the hopping model proposed in \cite{SH_bottom_up_2020}, we construct the matrix elements of $\hat{h} (\bm{r})$ and independent model parameters by means of the symmetry-adopted multipole basis.

\begin{figure*}[ht]
  \begin{center}
    \includegraphics[width=16cm]{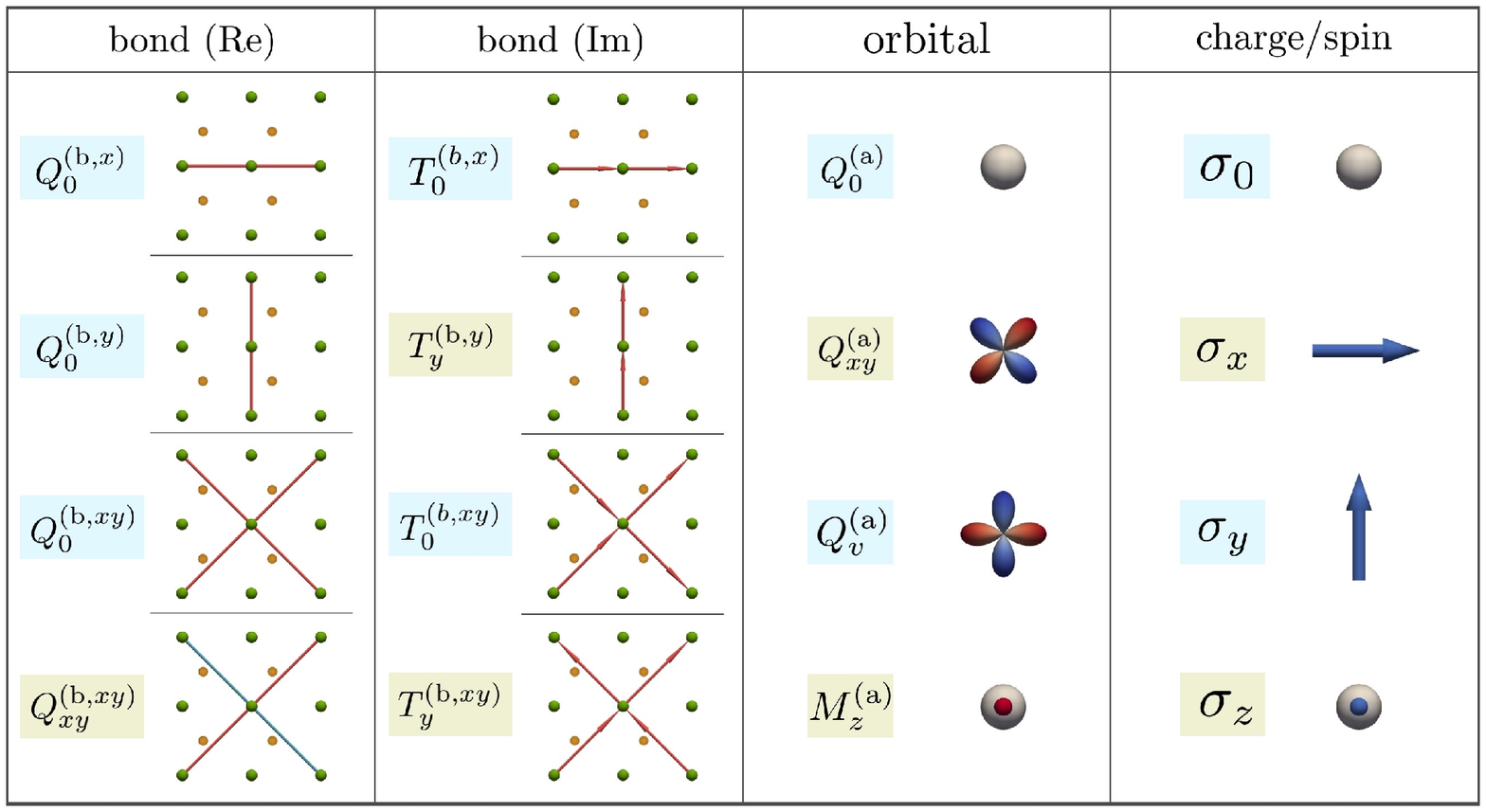}
  \end{center}
  \caption{
    Symmetry-adopted multipole basis for the bond, orbital, and spin degrees of freedom.
    The blue (green) cell denotes that its irreducible representation is $A'$ ($A''$) with even (odd) for the $xz$-mirror operation.
  }
  \label{fig_multipole_basis}
\end{figure*}

Let us first introduce the symmetry-adapted multipole basis~\cite{SH_MY_YY_HK_mul_2018, HK_RO_SH_comp_mul_2020, SH_bottom_up_2020}.
We consider the on-site/bond, orbital, and spin degrees of freedom separately.

The on-site energy is represented by the electric cluster multipole,
\begin{align}
   & Q_{0}^{({\rm c})} (\bm{0}) = 1,
  \label{eq_c}
\end{align}
where the superscript (c) denotes the cluster multipole (in this case, the size of the cluster is $1$).
On the other hand, the real and imaginary hoppings along $x$ and $y$ directions are represented by
\begin{align}
   & Q_{0}^{({\rm b}, x)} (\bm{a}_{1}) = Q_{0}^{({\rm b}, y)} (\bm{a}_{2}) = 1,
  \cr
   & T_{0}^{({\rm b}, x)} (\bm{a}_{1}) = T_{y}^{({\rm b}, y)} (\bm{a}_{2}) = i,
  \label{eq_bxby_Tx_Ty}
\end{align}
where the superscripts (b, $x$) and (b, $y$) denote the bond multipoles along $x$ and $y$ directions.
Note that from the symmetry point of view, the real and imaginary hoppings correspond to the electric charge ($Q$) and magnetic toroidal monopole ($T_{0}$) or dipole ($T_{y}$) at the bond center, respectively.
Similarly, the hoppings along the diagonal are given by
\begin{align}
   & Q_{0}^{({\rm b}, xy)} (\pm\bm{a}_{1}+\bm{a}_{2}) = 1, \cr
   & Q_{xy}^{({\rm b}, xy)} (\pm\bm{a}_{1}+\bm{a}_{2}) = \pm 1, \cr
   & T_{0}^{({\rm b}, xy)} (\bm{a}_{1} \pm \bm{a}_{2}) = i, \cr
   & T_{y}^{({\rm b}, xy)} (\pm\bm{a}_{1} + \bm{a}_{2}) = i,
  \label{eq_bxy_Ty}
\end{align}
where $\bm{a}_{1}$ and $\bm{a}_{2}$ are the unit vectors, whose lengths are set as unity.
Since the bond multipole basis in Eqs.~(\ref{eq_bxby_Tx_Ty}) and (\ref{eq_bxy_Ty}) are hermitian, the matrix elements for the opposite hopping direction are given by the relation in Eq.~(\ref{eq_O_hermite}).
The above multipole basis are shown schematically in the columns of ``bond'' in Fig.~\ref{fig_multipole_basis}.

In the orbital space, $\left\{\ket{p_{x}}, \ket{p_{y}}\right\}$, any matrix is described by the four spinless atomic multipoles as
\begin{align}
      & Q_{0}^{({\rm a})}
  = \begin{pmatrix} 1 & 0 \\ 0 & 1 \end{pmatrix}
  ,\quad
  Q_{xy}^{({\rm a})}
  = \begin{pmatrix} 0 & 1 \\ 1 & 0 \end{pmatrix}
  ,\quad
  Q_{v}^{({\rm a})}
  = \begin{pmatrix} 1 & 0  \\ 0 & -1 \end{pmatrix}
  ,
  \cr &
  M_{z}^{({\rm a})}
  = \begin{pmatrix} 0 & -i \\ i & 0 \end{pmatrix},
  \label{eq_a_b}
\end{align}
where the superscript (a) denotes the atomic multipole.
$M_{z}^{\rm (a)}$ represents the orbital magnetic dipole.
Similarly, in the spin sectors, $\left\{\uparrow, \downarrow\right\}$, any matrix is expressed by the Pauli and unit matrices as
\begin{align}
        & \sigma_{x}
  = \begin{pmatrix} 0 & 1 \\ 1 & 0 \end{pmatrix}
  ,\quad
  \sigma_{y}
  = \begin{pmatrix} 0 & -i \\ i & 0 \end{pmatrix}
  ,\quad
  \sigma_{z}
  = \begin{pmatrix} 1 & 0  \\ 0 & -1 \end{pmatrix}
  , \cr &
  \sigma_{0}
  = \begin{pmatrix} 1 & 0 \\ 0 & 1 \end{pmatrix}.
  \label{eq_pauli}
\end{align}
These multipoles are shown in the columns of ``orbital'' and ``charge/spin'' in Fig.~\ref{fig_multipole_basis}, respectively.

Since the Hamiltonian $\hat{h}$ is fully symmetric for all the symmetry operations, only the independent products of the cluster/bond, atomic orbital, and spin multipoles belonging to $A'$ irreducible representation contribute to $\hat{h}$.
Considering the multipole basis classified in $A'$ (blue) and $A''$ (green) irreducible representations in Fig.~\ref{fig_multipole_basis} and $A'\otimes A'=A''\otimes A''=A'$, $A'\otimes A''=A''$, we obtain
\begin{align}
      & \hat{h}
  =
  \hat{h}_{{\rm c}}^{} + \hat{h}_{\rm b}^{(x)} + \hat{h}_{\rm b}^{(y)} + \hat{h}_{\rm b}^{(xy)} + \hat{h}_{\rm SOC}^{},
  \cr & \quad
  \hat{h}_{{\rm c}}^{}
  =
  \epsilon_{1} Q_{0}^{({\rm c})} Q_{0}^{({\rm a})} \sigma_{0}
  +
  \epsilon_{2} Q_{0}^{({\rm c})} Q_{v}^{({\rm a})} \sigma_{0},
  \cr & \quad
  \hat{h}_{\rm b}^{(x)}
  =
  t_{x1} Q_{0}^{({\rm b}, x)} Q_{0}^{({\rm a})} \sigma_{0}
  +
  t_{x2} Q_{0}^{({\rm b}, x)} Q_{v}^{({\rm a})} \sigma_{0},
  \cr & \quad
  \hat{h}_{\rm b}^{(y)}
  =
  t_{y1} Q_{0}^{({\rm b}, y)} Q_{0}^{({\rm a})} \sigma_{0}
  +
  t_{y2} Q_{0}^{({\rm b}, y)} Q_{v}^{({\rm a})} \sigma_{0}
  \cr
      & \hspace{3cm}
  +
  t_{y3} T_{y}^{({\rm b}, y)} M_{z}^{({\rm a})} \sigma_{0}
  ,
  \cr & \quad
  \hat{h}_{\rm b}^{(xy)}
  =
  t_{1}' Q_{0}^{({\rm b}, xy)} Q_{0}^{({\rm a})} \sigma_{0}
  +
  t_{2}' Q_{0}^{({\rm b}, xy)} Q_{v}^{({\rm a})} \sigma_{0}
  \cr & \hspace{2cm}
  +
  t_{3}' Q_{xy}^{({\rm b}, xy)} Q_{xy}^{({\rm a})} \sigma_{0}
  +
  t_{4}' T_{y}^{({\rm b}, xy)} M_{z}^{({\rm a})} \sigma_{0},
  \cr & \quad
  \hat{h}_{\rm SOC}^{}
  =
  \lambda Q_{0}^{({\rm c})} M_{z}^{({\rm a})} \sigma_{z}
  ,
  \label{eq_H_snte}
\end{align}
where $\epsilon_{i}$ ($i=1,2$), $t_{xi}$ ($i=1,2$), $t_{yi}$ ($i=1,2,3$), and $t_{i}'$ ($i=1,2,3,4$) are the independent model parameters for the on-site energies and transfer integrals, respectively.
$\lambda$ represents the atomic SOC magnitude.
We neglect the spin-dependent hoppings for simplicity, as the following results are not altered qualitatively.
Note that $t_{y3}$ and $t_{4}'$ are proportional to the electric polarization~\cite{snte_kim_2019}, which vanish in the paraelectric phase.

In the momentum-space representation, Eq.~(\ref{eq_H_snte}) is expressed as
\begin{align}
      & \hat{h}_{\rm b}^{(x)}(\bm{k}) =
  t_{x1} Q_{0}^{({\rm b}, x)} (\bm{k}) Q_{0}^{({\rm a})} \sigma_{0}
  +
  t_{x2} Q_{0}^{({\rm b}, x)} (\bm{k}) Q_{v}^{({\rm a})} \sigma_{0},
  \cr & \hat{h}_{\rm b}^{(y)}(\bm{k}) =
  t_{y1} Q_{0}^{({\rm b}, y)} (\bm{k}) Q_{0}^{({\rm a})} \sigma_{0}
  +
  t_{y2} Q_{0}^{({\rm b}, y)} (\bm{k}) Q_{v}^{({\rm a})} \sigma_{0}
  \cr & \hspace{2cm}
  +
  t_{y3} T_{y}^{({\rm b}, y)} (\bm{k}) M_{z}^{({\rm a})} \sigma_{0}
  ,
  \cr & \hat{h}_{\rm b}^{(xy)} (\bm{k})
  =
  t_{1}' Q_{0}^{({\rm b}, xy)} (\bm{k}) Q_{0}^{({\rm a})} \sigma_{0}
  +
  t_{2}' Q_{0}^{({\rm b}, xy)} (\bm{k}) Q_{v}^{({\rm a})} \sigma_{0}
  \cr & \hspace{1cm}
  +
  t_{3}' Q_{xy}^{({\rm b}, xy)} (\bm{k}) Q_{xy}^{({\rm a})} \sigma_{0}
  +
  t_{4}' T_{y}^{({\rm b}, xy)} (\bm{k}) M_{z}^{({\rm a})} \sigma_{0},
  \label{eq_Hxy_k}
\end{align}
where the form factors are given by using Eq.~(\ref{eq_Ok}) as
\begin{align}
   & Q_{0}^{({\rm b}, x)} (\bm{k}) = 2 \cos (k_{x}), \quad Q_{0}^{({\rm b}, y)} (\bm{k}) = 2 \cos (k_{y}),  \cr
   & Q_{0}^{({\rm b}, xy)} (\bm{k}) =  4 \cos(k_{x}) \cos (k_{y}),                                           \cr
   & Q_{xy}^{({\rm b}, xy)} (\bm{k}) = -4 \sin(k_{x}) \sin (k_{y}),                                          \cr
   & T_{0}^{({\rm b}, x)} (\bm{k}) = 2 \sin (k_{x}), \quad  T_{y}^{({\rm b}, y)} (\bm{k}) = 2 \sin (k_{y}), \cr
   & T_{0}^{({\rm b}, xy)} (\bm{k}) = 4 \sin(k_{x}) \cos(k_{y}),                                             \cr
   & T_{y}^{({\rm b}, xy)} (\bm{k}) = 4 \cos(k_{x}) \sin(k_{y}).
\end{align}

The resultant Hamiltonian matrix is simply given by
\begin{align}
  \hat{h}(\bm{k})=\begin{pmatrix}
    h_{11}(\bm{k})     & h_{12}(\bm{k}) \\
    h_{12}^{*}(\bm{k}) & h_{22}(\bm{k})
  \end{pmatrix},
\end{align}
where
\begin{align}
      &
  h_{11}(\bm{k})=\left(\epsilon_{x} + p_{x1}c_{x} + p_{y1}c_{y} + p_{1}'c_{x}c_{y}\right) \sigma_{0},
  \cr &
  h_{12}(\bm{k})=(- i p_{y3} s_{y} + p_{3}' s_{x}s_{y} - i p_{4}' c_{x} s_{y})\sigma_{0} - i \lambda \sigma_{z},
  \cr &
  h_{22}(\bm{k})=\left(\epsilon_{y} + p_{x2}c_{x} + p_{y2}c_{y} + p_{2}'c_{x}c_{y}\right) \sigma_{0}.
  \label{eq_Hk_matrix}
\end{align}
Here, we use abbreviations, $c_{i} \equiv \cos(k_{i}),\, s_{i} \equiv\sin(k_{i}) \, (i = x, y)$ and
\begin{align}
   & \epsilon_{x} = \epsilon_{1} + \epsilon_{2},\,\,\, \epsilon_{y} = \epsilon_{1} - \epsilon_{2}, \cr
   & p_{x1} = 2(t_{x1} + t_{x2}),\,\,\, p_{x2} = 2(t_{x1} - t_{x2}),                \cr
   & p_{y1} = 2(t_{y1} + t_{y2}),\,\,\, p_{y2} = 2(t_{y1} - t_{y2}),\,\,\, p_{y3} = 2 t_{y3},                          \cr
   & p_{1}' = 4(t_{1}' + t_{2}'),\,\,\, p_{2}' = 4(t_{1}' - t_{2}'),\,\,\, p_{3}' = -4 t_{3}',\,\,\, p_{4}' = 4 t_{4}'. \cr
\end{align}

\subsection{Antisymmetric Spin and Orbital Splittings}
\label{sec:antisplit}

In the absence of the space-inversion symmetry, the antisymmetric spin and orbital splittings in the electronic band structure are expected.
The spin and orbital splittings are evaluated by using Eq.~(\ref{eq_g0a_k_1}).
The lowest-order contribution to the anti-symmetric spin splitting is found at $i = 2$ as
\begin{align}
  \Omega^{2}_{\sigma_{z}} (\bm{k})
      & =
  \mathrm{Tr} \left[\sigma_{z}\hat{h}^{2}(\bm{k})\right]
  \cr &
  = 8 \lambda
  \left( t_{y3} T_{y}^{({\rm b}, y)} (\bm{k}) + t_{4}' T_{y}^{({\rm b}, xy)} (\bm{k}) \right)
  \cr &
  = 16 \lambda
  \left(t_{y3} + 2 t_{4}' \cos \left(k_{x}\right)\right)
  \sin \left(k_{y}\right)
  \cr &
  \to \lambda \left(t_{y3} + 2 t_{4}'\right) k_{y}\quad (\bm{k}\to\bm{0}).
  \label{eq_gszk}
\end{align}
Similarly, the lowest-order contribution to the anti-symmetric orbital splitting is given by
\begin{align}
  \Omega^{2}_{M_{z}} (\bm{k})
      & =
  \mathrm{Tr}
  \left[ M_{z}^{({\rm a})} \hat{h}^{2}(\bm{k}) \right]
  \cr &
  =
  4 \left(
  t_{y3} T_{y}^{({\rm b}, y)} (\bm{k}) + t_{4}' T_{y}^{({\rm b}, xy)} (\bm{k})
  \right)
  \cr &
  =
  8 \left( t_{y3} + 2 t_{4}' \cos \left(k_{x}\right) \right)
  \sin \left(k_{y}\right)
  \cr &
  \to
  \left( t_{y3} + 2 t_{4}' \right) k_{y} \quad (\bm{k}\to\bm{0}).
  \label{eq_glzk}
\end{align}
These results indicate that the Rashba-type spin and orbital splittings appear in the $k_{y}$-direction, and degenerate at $k_{y} = 0$ because of the mirror symmetry in the $xz$ plane.
Equations~(\ref{eq_gszk}) and (\ref{eq_glzk}) clearly indicate that the anti-symmetric splitting requires finite $t_{y3}$ or $t_{4}'$ hoppings, which indeed become finite when the system enters in the ferroelectric phase.
Moreover, $\lambda$ is necessary (unnecessary) for the spin (orbital) splitting.
These results are consistent with those by the {\it ab initio} calculation~\cite{snte_kim_2019}, and are summarized in Table~\ref{tab_params}.

\begin{figure}[t]
  \begin{center}
    \includegraphics[width=9cm]{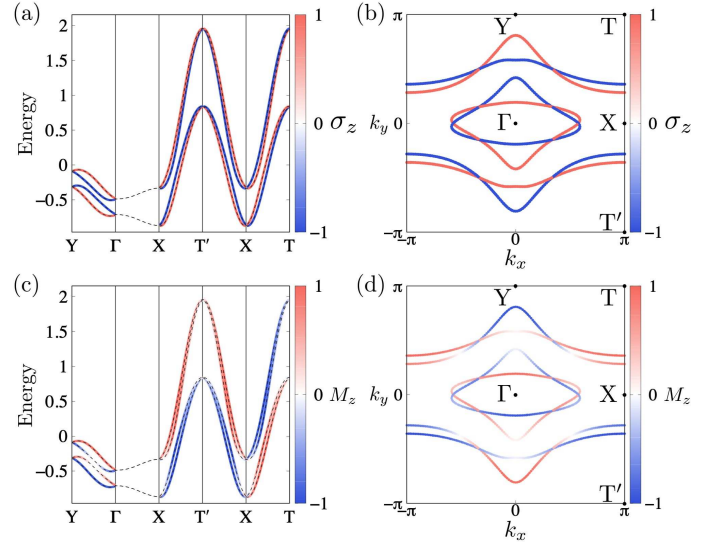}
  \end{center}
  \caption{
    Conduction band structure of the effective model, whose parameters are given in Eq.~(\ref{eq_params}).
    The color map of the $z$ component of (a) the spin and (c) orbital polarization along the high symmetry line in $\bm{k}$ space.
    The dashed lines represent the band dispersions.
    (b), (d) The isoenergy surfaces correspond to (a) and (c) at $\mu = -0.4$.
  }
  \label{band_disp}
\end{figure}

We show the electronic band structure of this model in Figs.~\ref{band_disp}(a)-(d), where the hopping parameters and the strength of the SOC are set as
\begin{align}
      & \epsilon_{1} = 0.0, \qquad\, \epsilon_{2} = 0.05,
  \cr &
  t_{x1} = -0.2, \quad t_{x2} = -0.05,
  \cr &
  t_{y1} = -0.3, \quad t_{y2} = -0.1, \quad t_{y3} = -0.05,
  \cr &
  t_{1}' = 0.1, \qquad\, t_{2}' = 0.05, \quad\,\,\,\,\, t_{3}' = 0.05, \quad t_{4}' = 0.05,
  \cr &
  \lambda = 0.1.
  \label{eq_params}
\end{align}
Note that these parameters are chosen so as to reproduce the conduction band edge near the X point given by the {\it ab initio} calculation~\cite{snte_kim_2019}.
As shown in Figs.~\ref{band_disp}(a)-(d), the antisymmetric spin and orbital splittings in the $k_{y}$-direction appear, while there is no spin and orbital splittings at $k_{y} = 0$, which are consistent with Eqs.~(\ref{eq_gszk}) and (\ref{eq_glzk}), respectively.
In addition, we have confirmed that the spin and orbital splittings disappear when the essential parameters $t_{y3}$ and $t_{4}'$ are set to 
zero.

\begin{table}[t]
  \caption{
    Essential parameters for the band splitting and responses indicated by the checkmark (\checkmark).
  }
  \label{tab_params}
  \begin{center}
    \begin{tabular}{lcrr}
      \hline \hline
                                     & $t_{y3}$ or $t_{4}'$ & $\,\,\, t_{3}'$ & $\,\,\,\lambda$ \\ \hline
      spin splitting                 & \checkmark           &                 & \checkmark      \\
      orbital splitting              & \checkmark           &                 &                 \\ \hline
      Nonlinear Hall effect          & \checkmark           & \checkmark      &                 \\
      orbital magneto-current effect & \checkmark           &                 &                 \\
      \hline \hline
    \end{tabular}
  \end{center}
\end{table}

\subsection{Nonlinear Hall Effect}
\label{sec:nlhesnte}

Next, we elucidate the essential parameters and the microscopic picture of the nonlinear Hall effect. 
Since the system satisfies the symmetry, $(\mathcal{P}, \mathcal{T}, \mathcal{PT}) = (\times, \bigcirc, \times)$, we only need to evaluate the BCD term of the second-order conductivity, $\sigma_{x;y,y}^{\rm BCD}$, as shown in Table~\ref{tab_responses}.

The lowest-order contribution to $\sigma_{x;y,y}^{\rm BCD}$ arises at $(i,j,k) = (0,1,0)$ in the last term of Eq.~(\ref{eq_cond_2_bcd_k}) as
\begin{align}
      & \mathrm{Im}\left[\Gamma^{0,1,0}_{x; y,y}\right]
  =
  \sum_{\bm{k}} \mathrm{Tr}
  \left[\hat{v}_{x}(\bm{k})\hat{v}_{y}(\bm{k})\hat{h}(\bm{k})\hat{v}_{y}(\bm{k})\right]
  \cr & \quad
  =
  -128
  t_{3}'
  \left(
  t_{2}' t_{y3} t_{y1}
  + t_{y2} t_{y3} t_{1}'
  + 2 t_{y2} t_{4}' t_{y1}
  + 4 t_{2}' t_{4}' t_{1}'
  \right),
  \cr &
  \label{eq_bcd_010_k}
\end{align}
which indicates that the $\lambda$ is unnecessary as pointed out in the previous study~\cite{snte_kim_2019}.
Furthermore, it indicates that not only $t_{y3}$ or $t_{4}'$ but also the diagonal hopping $t_{3}'$ involving the electric quadrupole $Q_{xy}^{(a)}$ is essential.
Indeed, the higher-order terms in Eq.~(\ref{eq_cond_2_bcd_k}) always contain $t_{y3}$ or $t_{4}'$, and are proportional to $t_{3}'$.
As a result,  $\sigma_{x;y,y}^{\rm BCD}$ is expressed in the form,
\begin{align}
  \sigma_{x;y,y}^{\rm BCD} = i t_{3}' \left\{ t_{y3} F (\bm{t}) + t_{4}' F' (\bm{t}) \right\},
\end{align}
where $F (\bm{t})$ and $F' (\bm{t})$ denote functions of the model parameters.
The essential parameters for the nonlinear Hall effect are summarized in Table~\ref{tab_params}.

\begin{figure*}[t]
  \begin{center}
    \includegraphics[width=17.0cm]{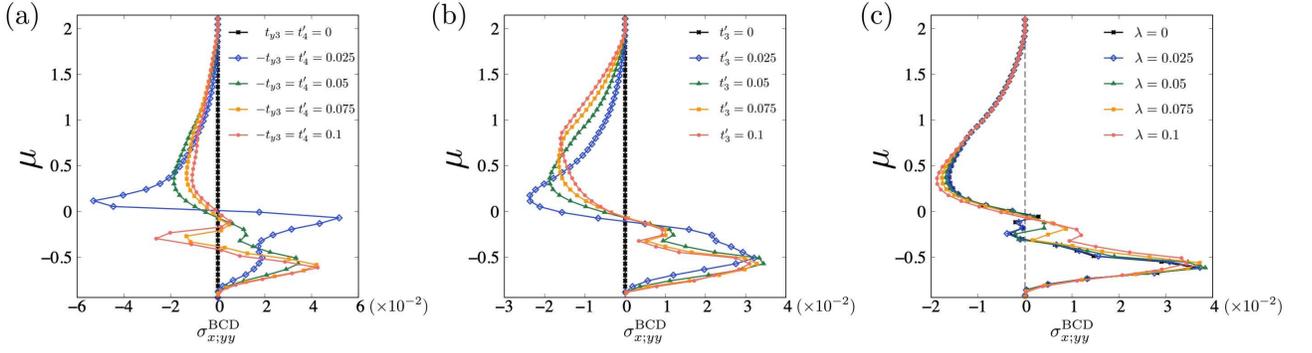}
  \end{center}
  \caption{
  Chemical potential $\mu$ dependencies with or without the essential parameters of the BCD term in the second-order conductivity $\sigma_{x;y,y}^{\rm BCD}$ evaluated by Eq.~(\ref{eq_bcd_kubo}) at $T = 0.01$ and $\hbar = e = \gamma = 1$.
  The model parameters correspond to Eq.~(\ref{eq_params}), and the $\bm{k}$ mesh $N$ is chosen as $256^{2}$.
  (a) $-t_{y3} = t_{4}'$, (b) $t_{3}'$, and (c) $\lambda$ dependencies.
  }
  \label{fig_bcd_cond}
\end{figure*}

We confirm the above results by direct numerical evaluation.
$\sigma_{x;y,y}^{\rm BCD}$ obtained by Eq.~(\ref{eq_bcd_kubo}), which contains all orders of contributions in Eq.~(\ref{eq_cond_2_bcd_k}), is shown in Fig.~\ref{fig_bcd_cond}.
The results clearly show that when the essential parameters are set to zero, $\sigma_{x;y,y}^{\rm BCD}$ vanishes irrespective of the value of $\mu$, whereas it remains finite without $\lambda$.
In this way, the present analysis method can systematically identify the essential parameters in the response tensors.

\begin{figure*}[ht]
  \begin{center}
    \includegraphics[width=16cm]{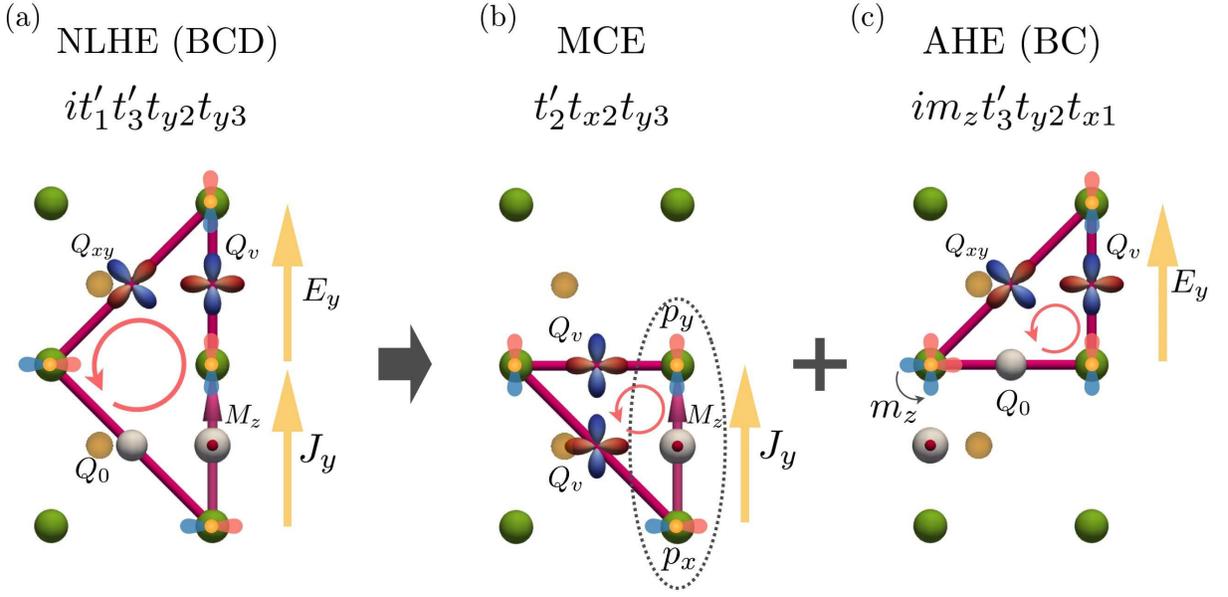}
  \end{center}
  \caption{
  Microscopic picture of the nonlinear Hall effect.
  (a) The lowest-order contribution to $\sigma_{x;y,y}^{\rm BCD}$.
  This process is understood by the following two processes.
  (b) The orbital magneto-current effect: $M_{z}^{({\rm a})}$ is induced by the electric current $J_{y}$ with the hoppings activated by the ferroelectricity.
  The dashed oval indicates the microscopic origin of the orbital magneto-current effect.
  (c) The anomalous Hall effect: the Hall conductivity arises in the presence of the induced orbital magnetic moment $m_{z}$.
  }
  \label{fig_nlahe}
\end{figure*}

By analyzing the lowest-order contribution Eq.~(\ref{eq_bcd_010_k}) in the real-space representation, we obtain microscopic picture of the nonlinear Hall effect.
This is given by applying the relation Eq.~(\ref{eq_Ok}) to $\hat{v}_{x}(\bm{k})$, $\hat{v}_{y}(\bm{k})$, and $\hat{h}(\bm{k})$ in Eq.~(\ref{eq_bcd_010_k}).
In the real-space representation, the nonzero trace in Eqs.~(\ref{eq_gam_0_a})-(\ref{eq_gam_2_ab}) corresponds to a closed loop of the bonds of the operators.
Then, $\Omega^{0,1,0}_{x; y,y}(C)$ corresponding to Eq.~(\ref{eq_bcd_010_k}) is given by
\begin{align}
   & \Omega^{0,1,0}_{x; y,y}(C)=
  \mathrm{Tr}
  \left[
    \hat{v}_{x}(\bm{a}_{1}-\bm{a}_{2})
    \hat{v}_{y}(-\bm{a}_{1}-\bm{a}_{2})
    \hat{h} (\bm{a}_{2})
    v_{y}(\bm{a}_{2})
    \right],
  \label{eq_bcd_010_r}
\end{align}
where the closed loop, $C:\{\bm{a}_{2}, \bm{a}_{2}, -(\bm{a}_{1} + \bm{a}_{2}), \bm{a}_{1} - \bm{a}_{2}\}$ is shown in Fig.~\ref{fig_nlahe}(a).

By comparing Eq.~(\ref{eq_bcd_010_r}) with Eq.~(\ref{eq_bcd_010_k}), we identify the building blocks of the contributions as
\begin{align}
  \begin{split}
    & \hat{v}_{x}(\bm{a}_{1}-\bm{a}_{2}) \to
    -i t_{1}' Q_{0}^{({\rm a})}, \\
    & \hat{v}_{y}(-\bm{a}_{1}-\bm{a}_{2}) \to
    i t_{3}' Q_{xy}^{({\rm a})}, \\
    & \hat{h} (\bm{a}_{2}) \to
    t_{y2} Q_{v}^{({\rm a})}, \\
    & \hat{v}_{y}(\bm{a}_{2}) \to
    t_{y3}  M_{z}^{({\rm a})},
  \end{split}
\end{align}
where we have used Eq.~(\ref{eq_velo_real}).
Indeed, by substituting them into Eq.~(\ref{eq_bcd_010_r}), we obtain
\begin{align}
  \Omega^{0,1,0}_{x; y,y}(C)
      & \to t_{1}' t_{3}' t_{y2} t_{y3}
  \,
  \mathrm{Tr}
  \left[
  Q_{0}^{({\rm a})}Q_{xy}^{({\rm a})}Q_{v}^{({\rm a})}M_{z}^{({\rm a})}
  \right]
  \cr &
  \propto i\, t_{1}' t_{3}' t_{y2} t_{y3}.
  \label{eq_bcd_cl}
\end{align}
The schematic picture of Eq.~(\ref{eq_bcd_cl}) is shown in Fig.~\ref{fig_nlahe}(a).
Similarly, other closed loops give the rest of terms in Eq.~(\ref{eq_bcd_010_k}).

Based on Eq.~(\ref{eq_bcd_cl}), the nonlinear Hall effect can be interpreted by the subsequent two processes: the orbital magneto-current effect and the linear anomalous Hall effect triggered by the induced orbital magnetization.
The linear orbital magneto-current tensor is defined as~\footnote{(Supplemental Material) Derivation of the explicit form of the linear and second-order electric-field/current induced response tensors in the velocity gauge is given.}
\begin{align}
  \braket{M_{z}^{({\rm a})}} = \alpha^{({\rm J})}_{z;y} E_{y}.
\end{align}
The superscript (J) indicates the dissipative part.
The lowest-order contribution to $\alpha^{({\rm J})}_{z; y}$ arises at $(i, j) = (2, 0)$ as
\begin{align}
      & \mathrm{Re}\left[\Gamma^{2,0}_{z;y}\right]
  =
  \sum_{\bm{k}}\mathrm{Tr}
  \left[M_{z}^{({\rm a})} \hat{h}^{2}(\bm{k}) \hat{v}_{y}(\bm{k}) \right]
  \cr & \quad
  =
  16\left[
    t_{y3}\left( \epsilon_{2} t_{y2} + 2 t_{x2} t_{2}' \right)
    +
    2 t_{4}'\left( \epsilon_{2} t_{2}' + t_{x2} t_{y2} \right)
    \right].
  \cr &
  \label{eq_aJlz_20_k}
\end{align}
Thus, the orbital magneto-current effect occurs in the presence of the hoppings $t_{y3}$ or $t_{4}'$, as depicted in Fig.~\ref{fig_nlahe}(b).

In the real-space representation, considering a closed loop, $C : \{\bm{a}_{2}, -\bm{a}_{1}, \bm{a}_{1} - \bm{a}_{2}, \bm{0}\}$, we obtain
\begin{align}
  \Omega^{2,0}_{z; y} (C)
  =
  \mathrm{tr}
  \left[
  M_{z}^{({\rm a})} \hat{h} (\bm{a}_{1}-\bm{a}_{2}) \hat{h} (-\bm{a}_{1}) \hat{v}_{y}(\bm{a}_{2})
  \right].
  \label{eq_aJlz_20_r}
\end{align}
By the correspondences,
\begin{align}
  \begin{split}
    & \hat{h} (\bm{a}_{1}-\bm{a}_{2}) \to t_{2}' Q_{v}^{({\rm a})}, \\
    & \hat{h} (-\bm{a}_{1}) \to t_{x2} Q_{v}^{({\rm a})}, \\
    & \hat{v}_{y}(\bm{a}_{2}) \to t_{y3}  M_{z}^{({\rm a})},
  \end{split}
\end{align}
we find
\begin{align}
  \Omega^{2,0}_{z; y} (C)
   & \to
  t_{2}' t_{x2} t_{y3}
  \mathrm{tr}
  \left[
  M_{z}^{({\rm a})} Q_{v}^{({\rm a})} Q_{v}^{({\rm a})} M_{z}^{({\rm a})}
  \right]
  \propto t_{2}' t_{x2} t_{y3}.
  \label{eq_aJlz_cl}
\end{align}
The essential parameters for the magneto-current effect are summarized in Table~\ref{tab_params}.

Next, we analyze the linear anomalous Hall effect that arises in the presence of the above induced orbital magnetization.
By adding the mean-field term $\hat{H}_{\rm MF} = m_{z} M_{z}^{({\rm a})}$ to the Hamiltonian as the induced orbital magnetization, we calculate the BC term of the linear conductivity $\sigma^{\rm BC}_{x;y} (m_{z})$ in Eq.~(\ref{eq_cond_1_bc_k}).
The lowest-order contribution to $\sigma^{\rm BC}_{x;y} (m_{z})$ arises at $(i,j) = (2, 0)$ as
\begin{align}
  \mathrm{Im}\left[\Gamma^{2,0}_{x;y}\right]
      & =
  \sum_{\bm{k}}\mathrm{Tr}
  \left[\hat{v}_{x}(\bm{k})\hat{h}^{2}(\bm{k})\hat{v}_{y}(\bm{k})\right]
  \cr &
  =
  32
  \, m_{z} t_{3}' \left(t_{x1} t_{y2} - t_{y1} t_{x2}\right).
  \label{eq_bc_20_k}
\end{align}
The result indicates that the diagonal hopping $t_{3}'$ is the essential parameter for the anomalous Hall effect under the mean field $m_{z}$.

\begin{figure}[t]
  \begin{center}
    \includegraphics[width=8cm]{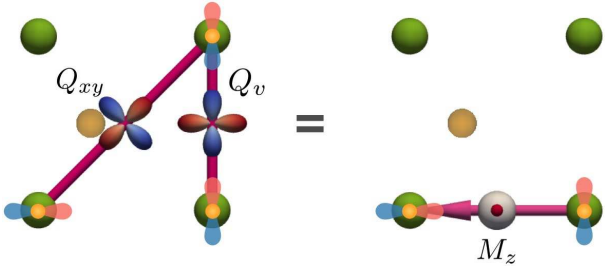}
  \end{center}
  \caption{
  Effective hopping with $M_{z}^{({\rm a})}$ along $x$ direction, which arises from $t_{y2}$ and $t_{3}'$.
  }
  \label{fig_hidden_mul}
\end{figure}

In the real-space representation, considering a closed loop, $C : \{\bm{a}_{2}, -\bm{a}_{1}-\bm{a}_{2}, \bm{0}, \bm{a}_{1}\}$ as shown in Fig.~\ref{fig_nlahe}(c), we obtain
\begin{align}
  \Omega^{20}_{x; y} (C)
  =
  \mathrm{Tr}
  \left[
    \hat{v}_{x} (\bm{a}_{1})\hat{h} (\bm{0})\hat{h} (-\bm{a}_{1}-\bm{a}_{2})\hat{v}_{y}(\bm{a}_{2})
    \right].
  \label{eq_bc_20_r}
\end{align}
By the correspondences,
\begin{align}
  \begin{split}
    & \hat{v}_{x} (\bm{a}_{1}) \to -i t_{x1} Q_{0}^{({\rm a})}, \\
    & \hat{h} (\bm{0}) \to m_{z} M_{z}^{({\rm a})}, \\
    & \hat{h} (-\bm{a}_{1}-\bm{a}_{2}) \to t_{3}' Q_{xy}^{({\rm a})}, \\
    & \hat{v}_{y}(\bm{a}_{2}) \to -i t_{y2} Q_{v}^{({\rm a})},
  \end{split}
\end{align}
Eq.~(\ref{eq_bc_20_r}) reads
\begin{align}
  \Omega^{20}_{x; y} (C)
      & \to -m_{z} t_{3}' t_{y2} t_{x1}
  \,
  \mathrm{Tr}
  \left[Q_{0}^{({\rm a})}M_{z}^{({\rm a})}Q_{xy}^{({\rm a})}Q_{v}^{({\rm a})}\right]
  \cr &
  \propto i\, m_{z} t_{3}' t_{y2} t_{x1}.
  \label{eq_bc_cl}
\end{align}
As depicted in Fig.~\ref{fig_hidden_mul}, the combination of $t_{y2}$ and $t_{3}'$ generates the effective hopping accompanied by the orbital magnetic dipole moment, $Q_{xy}^{({\rm a})} Q_{v}^{({\rm a})} = i M_{z}^{({\rm a})}$ along $x$ direction.
This term itself is forbidden by the $xz$-mirror symmetry, hence it does not appear in the Hamiltonian.
However, in the higher-order process, the effective coupling between the hopping and the hidden orbital magnetization arises as shown in Fig.~\ref{fig_nlahe}.

\section{Summary}
\label{sec:summary}

In summary, we have proposed a systematic analysis method for identifying essential parameters in various types of response tensors.
By analyzing the power series of the Hamiltonian matrix of a given model, we extract the essential parameters in the response tensors.
Analyzing the low-order contributions in the real-space representation, we also obtain microscopic picture of the response.
The summary of the relevant explicit expressions to evaluate the essential parameters for thermal average, and linear and second-order response tensors are given in Table~\ref{tab_responses}.

We have demonstrated our method by analyzing the nonlinear Hall effect in the ferroelectric SnTe monolayer, and revealed that the second-neighbor diagonal hopping corresponding to the cluster electric quadrupole is the essential parameter for the nonlinear Hall effect in the model, whereas the atomic spin-orbit coupling is not.
Considering the microscopic picture in the real space, we also find that the nonlinear Hall effect can be regarded as a combination of two processes: the current-induced orbital magnetization and the linear anomalous Hall effect triggered by the induced orbital magnetization.

In this way, the present method is useful to identify the essential parameters in response tensors and is quite compatible with computational analysis.
It unveils hidden couplings among the electron hoppings, SOC, and order parameters, and promote further investigation on the anomalous responses.

\begin{acknowledgments}
  We would like to thank Megumi Yatsushiro, Satoru Hayami, and Yukitoshi Motome for fruitful discussions.
  This work was supported by JSPS KAKENHI Grants Numbers JP19K03752, JPJP20J21838, and JP21H01031.
  A part of numerical and symbolic calculations was performed in the supercomputing systems in the MAterial science Supercomputing system for Advanced MUlti-scale simulations towards NExt-generation-Institute for Materials Research (MASAMUNE-IMR) of the Center for Computational Materials Science, Institute for Materials Research, Tohoku University.
\end{acknowledgments}

\appendix
\section{Keldysh Formalism}
\label{sec:kel}

\begin{figure}[b]
  \begin{center}
    \includegraphics[width=6cm]{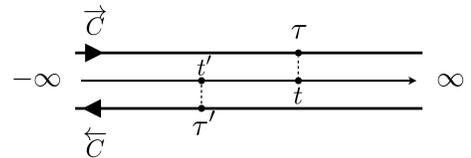}
  \end{center}
  \caption[]{
    Keldysh contour $C = \overrightarrow{C} + \overleftarrow{C}$.
    The forward (backward) contour $\overrightarrow{C}\, (\overleftarrow{C})$ goes from $-\infty\, (\infty)$ to $\infty\, (-\infty)$ in time.
    For example, $t > t'$ on the real-time axis, but $\tau \overset{C}{<} \tau'$ on the Keldysh contour $C$.
  }
  \label{fig_kel_contour}
\end{figure}

In this appendix, we give a short summary of the Keldysh formalism~\cite{Keldysh_1964}.
Let us begin with the time-dependent Hamiltonian $\mathcal{H}(t) = \mathcal{H}_{0} + \mathcal{H}_{\rm ext} (t)$, where $\mathcal{H}_{0}$ is the target system and $\mathcal{H}_{\rm ext}$ is the perturbation of the external fields.
In the Keldysh formalism, the non-equilibrium Green's function is introduced to use Wick's theorem for a finite temperature system or a system in a non-equilibrium state~\cite{jishi_2013}.
In order to define the non-equilibrium Green's function, we first introduce the Keldysh contour $C = \overrightarrow{C}+\overleftarrow{C}$ as shown in Fig.~\ref{fig_kel_contour}.
The contour-ordering operator for fermions along $C$ is defined by
\begin{align}
  T_{C}\left[A(\tau) B(\tau')\right]
  =\begin{cases}
    A(\tau) B(\tau')   & \tau \overset{C}{>} \tau', \\
    - B(\tau') A(\tau) & \tau'\overset{C}{>} \tau.
  \end{cases}
\end{align}
Using the time-ordering operator $T_{C}$, the non-equilibrium Green's function is given by
\begin{align}
  i G^{c}_{ij}(\tau , \tau') & = \Braket{T_{C} \left[ c^{\rm H}_{i}(\tau) c^{\dagger, {\rm H}}_{j}(\tau')\right]}_{0} \cr
                             & = \Braket{T_{C} \left[ S^{C}_{\rm ext} c^{\rm I}_{i}(\tau) c^{\dagger, {\rm I}}_{j}(\tau') \right]}_{0},
  \label{eq_Gc_interaction}
\end{align}
where the superscript H (I) represents that the operator is in the Heisenberg (interaction) representation, and $\braket{\cdots}_{0} = \mathrm{Tr}[ e^{- \beta (\mathcal{H}_{0} - \mu N)} \cdots ] / \mathrm{Tr}[ e^{- \beta (\mathcal{H}_{0} - \mu N)} ]$.
$S^{C}_{\rm ext}$ is the $S$-matrix:
\begin{align}
   & S^{C}_{\rm ext} = \exp \left(\int_{C} d \tau_{1}  V_{\rm I}(\tau)\right),
   & V(t) = \frac{1}{i \hbar} \mathcal{H}_{\rm ext}(t).
\end{align}
Expanding $S^{C}_{\rm ext}$ with respect to $V$ and applying Wick's theorem to Eq.~(\ref{eq_Gc_interaction}), $G^{c}$ can be expressed in the form of the Dyson's equation as
\begin{align}
  i \hat{G}^{c}(\tau, \tau')
   & =
  i \hat{\mathcal{G}}^{c}(\tau,\tau') + \int_{C} d\tau_{1} i \hat{\mathcal{G}}^{c}(\tau, \tau_{1}) \hat{V}(\tau_{1}) i \hat{G}^{c} (\tau_{1}, \tau'),
  \label{eq_Gc_dyson}
\end{align}
where $\mathcal{G}^{c}$ represents the unperturbed non-equilibrium Green's function.

To calculate various physical quantities, we introduce the real-time Green's functions.
Depending on whether $\tau, \tau'$ are on $\overrightarrow{C}$ or $\overleftarrow{C}$, $G^{c}$ is projected onto lesser ($G^{<}$), greater ($G^{>}$), time-ordered ($G^{T}$), and anti-time-ordered ($G^{\tilde{T}}$) Green's functions as
\begin{align}
   & i G^{<}_{ij} \left(t, t'\right)
  = -\braket{c_{j}^{\dagger, {\rm H}}(t') c_{i}^{\rm H}(t)}_{0},
  \, \quad (\tau \in \overrightarrow{C},  \tau' \in \overleftarrow{C}),
  \label{eq_G_lesser}                        \\
   & i G^{>}_{ij} \left(t, t'\right)
  = \braket{c_{i}^{\rm H}(t) c_{j}^{\dagger, {\rm H}}(t')}_{0},
  \quad (\tau' \in \overrightarrow{C},  \tau \in \overleftarrow{C}),
  \label{eq_G_greater}                       \\
   & i G^{T}_{ij} \left(t, t'\right)
  = \Braket{T_{\overrightarrow{C}} \left[
  c_{i}^{\rm H}(t) c_{j}^{\dagger, {\rm H}}(t') \right]}_{0},
  \, \quad (\tau, \tau' \in \overrightarrow{C}),
  \label{eq_G_T}                             \\
   & i G^{\tilde{T}}_{ij} \left(t, t'\right)
  = \Braket{T_{\overleftarrow{C}} \left[c_{i}^{\rm H}(t) c_{j}^{\dagger, {\rm H}}(t') \right]}_{0},
  \, \quad (\tau, \tau' \in \overleftarrow{C}).
  \label{eq_G_antiT}
\end{align}
By using these Green's functions, the retarded and advanced Green's functions are expressed as
\begin{align}
   & \hat{G}^{\rm R}(t,t') = \hat{G}^{T}(t,t') - \hat{G}^{<}(t,t'),
  \label{eq_G_R}                                                              \\
   & \hat{G}^{\rm A}(t,t') = - \hat{G}^{\tilde{T}}(t,t') + \hat{G}^{<}(t,t').
  \label{eq_G_A}
\end{align}

Among the real-time Green's functions, the lesser Green's function $G^{<}$ is significant as it is directly related to observables.
Using $G^{<}$, the ensemble average of arbitrary operator $\hat{O}$ can be expressed as
\begin{align}
  O(t) \equiv \braket{\hat{O}^{\rm H}(t)}
  = -i\,\mathrm{Tr}\left[\hat{O}(t) \hat{G}^{<} (t, t)\right].
\end{align}
Although $\hat{G}^{<}$ solely cannot be expanded in $\hat{V}$, Langreth rules~\cite{jishi_2013} give the perturbation expansion of $\hat{G}^{<}$ as
\begin{align}
      & i \hat{G}^{<}(t, t') = \sum_{n} i \hat{G}^{< (n)}(t, t')
  ,                                                                    \\
      & i \hat{G}^{< (n)}(t, t')
  =
  \sum_{\set{C}} \left(\prod_{j = 1}^{n} \int dt_{j}\right)
  i \hat{\mathcal{G}}^{C_{1}}(t, t_{1}) \hat{V}(t_{1})
  \cr & \qquad\times i \hat{\mathcal{G}}^{C_{2}} (t_{1}, t_{2}) \cdots
  i \hat{\mathcal{G}}^{C_{n}} (t_{n-1}, t_{n}) \hat{V} (t_{n}) i \hat{\mathcal{G}}^{C_{n+1}}(t_{n}, t'),
\end{align}
where $\hat{\mathcal{G}}^{C_{i}}$ represents any of unperturbed retarded ($\hat{\mathcal{G}}^{\rm R}$), advanced ($\hat{\mathcal{G}}^{\rm A}$), and lesser ($\hat{\mathcal{G}}^{<}$) Green's functions.
The summation $\sum_{\set{C}}$ means that for given $k$, $(\hat{\mathcal{G}}^{C_{1}}, \ldots, \hat{\mathcal{G}}^{C_{k-1}})$, $\hat{\mathcal{G}}^{C_{k}}$, and $(\hat{\mathcal{G}}^{C_{k+1}}, \ldots, \hat{\mathcal{G}}^{C_{n+1}})$ are replaced with $\hat{\mathcal{G}}^{\rm R}$, $\hat{\mathcal{G}}^{<}$, and $\hat{\mathcal{G}}^{\rm A}$, respectively, where $k$ runs over from $1$ to $n+1$.
The Fourier transform of $\hat{G}^{< (n)}(t, t'=t)$ is given by
\begin{widetext}
  \begin{multline}
    i\hat{G}^{< (n)}(\omega)
    =
    \left(\prod_{j = 1}^{2 n + 1}
    \int_{\omega_{c_{j}}}
    \right) (2 \pi)^{n + 1}
    \delta(\omega + \omega_{c_{2n+1}} - \omega_{c_{1}})
    \left(\prod_{k = 0}^{n-1}
    \delta(\omega_{c_{2k+1}} - \omega_{c_{2k+2}} - \omega_{c_{2k+3}})\right)
    \\ \times
    \sum_{\left\{C\right\}}
    \left[
    i\hat{\mathcal{G}}^{C_{1}}(\omega_{c_{1}})
    \left(\prod_{l = 1}^{n}
    \hat{V}(\omega_{c_{2l}})
    i\hat{\mathcal{G}}^{C_{l+1}}(\omega_{c_{2l+1}})\right)
    \right],
    \label{eq_G_n}
  \end{multline}
  where $\int_{\omega} \equiv \int_{- \infty}^{\infty} d\omega/2\pi$.
  The explicit expressions of $\hat{G}^{< (n)}(\omega)$ up to $n = 2$ are give by
  \begin{align}
     & i \hat{G}^{< (0)}(\omega)
    = 2\pi i \delta(\omega)
    \int_{\omega_{c_{1}}}
    \hat{\mathcal{G}}^{<}\left(\omega_{c_{1}}\right),
    \label{eq_G_0}
  \end{align}
  \begin{multline}
    i\hat{G}^{< (1)}\left(\omega\right)
    =-(2\pi)^{2}
    \int_{\omega_{c_{1}\cdots c_{3}}}
    \delta(\omega_{c_{1}} - \omega_{c_{2}} - \omega_{c_{3}})
    \delta(\omega + \omega_{c_{3}} - \omega_{c_{1}})
    \\
    \times
    \left[
      \hat{\mathcal{G}}^{<}\left(\omega_{c_{1}}\right)
      \hat{V}^{}(\omega_{c_{2}})
      \hat{\mathcal{G}}^{\rm A}\left(\omega_{c_{3}}\right)
      +
      \hat{\mathcal{G}}^{\rm R}\left(\omega_{c_{1}}\right)
      \hat{V}^{}(\omega_{c_{2}})
      \hat{\mathcal{G}}^{<}\left(\omega_{c_{3}}\right)
      \right],
    \label{eq_G_1}
  \end{multline}
  and
  \begin{multline}
    i\hat{G}^{< (2)}\left(\omega\right)
    =-(2\pi)^{3}i
    \int_{\omega_{c_{1}\cdots c_{5}}}
    \delta(\omega_{c_{1}} - \omega_{c_{2}} - \omega_{c_{3}})
    \delta(\omega_{c_{3}} - \omega_{c_{4}} - \omega_{c_{5}})
    \delta(\omega + \omega_{c_{5}} - \omega_{c_{1}})
    \\
    \times
    \left[
      \hat{\mathcal{G}}^{<}\left(\omega_{c_{1}}\right)
      \hat{V}^{}(\omega_{c_{2}})
      \hat{\mathcal{G}}^{\rm A}\left(\omega_{c_{3}}\right)
      \hat{V}^{}(\omega_{c_{4}})
      \hat{\mathcal{G}}^{\rm A}\left(\omega_{c_{5}}\right)
      +
      \hat{\mathcal{G}}^{\rm R}\left(\omega_{c_{1}}\right)
      \hat{V}^{}(\omega_{c_{2}})
      \hat{\mathcal{G}}^{<}\left(\omega_{c_{3}}\right)
      \hat{V}^{}(\omega_{c_{4}})
      \hat{\mathcal{G}}^{\rm A}\left(\omega_{c_{5}}\right)
      \right.\\
      \left.
      +
      \hat{\mathcal{G}}^{\rm R}\left(\omega_{c_{1}}\right)
      \hat{V}^{}(\omega_{c_{2}})
      \hat{\mathcal{G}}^{\rm R}\left(\omega_{c_{3}}\right)
      \hat{V}^{}(\omega_{c_{4}})
      \hat{\mathcal{G}}^{<}\left(\omega_{c_{5}}\right)
      \right],
    \label{eq_G_2}
  \end{multline}
  where the explicit expressions of $\hat{\mathcal{G}}^{\rm R}, \hat{\mathcal{G}}^{\rm A},$ and $\hat{\mathcal{G}}^{<}$ are given by Eqs.~(\ref{eq_g_lesser}) and (\ref{eq_G_RA}).
\end{widetext}

\section{Derivation of NonLinear Conductivity}
\label{sec:nlc}

In this appendix, the outline of the derivation of the $n$-th order conductivity in the velocity gauge is given.
Let us begin with the ensemble average of the $k$-th order contribution of the current operator in the velocity gauge, Eq.~(\ref{eq_j_om}):
\begin{align}
   & j_{\mu}^{(k)} (\omega) = \sum_{l = 0} j_{\mu}^{(k)} (\omega; l), \cr & \quad
  j_{\mu}^{(k)} (\omega; l) = -i \int_{\omega_{a}} \mathrm{Tr}\left[\hat{j}_{\mu} (\omega; l) \hat{G}^{< (k)} (\omega - \omega_{a})\right].\quad\quad
\end{align}
\begin{widetext}
  Then, $j_{\mu}^{(k)} (\omega; l)$ is further expanded by substituting Eq.~(\ref{eq_B_om_n}) to $\hat{B}$ in $\hat{G}^{< (k)}$ as
  \begin{align}
        & j_{\mu}^{(k)} (\omega; l) = \sum_{m = 0} j_{\mu}^{(k)} (\omega; l, m), \cr
        & \quad j_{\mu}^{(k)} (\omega; l, m)
    =
    i(2 \pi)^{2k+2}\frac{e^{k+l+m+1}}{V\hbar^{k}\, l! }
    \sum_{m_{1}+\cdots+m_{k}}^{=m}
    \frac{1}{(m_{1}+1)!}\cdots\frac{1}{(m_{k}+1)!}
    \int_{\omega_{a}}
    \left(\prod_{j = 1}^{2 k + 1} \int_{\omega_{c_{j}}}\right)
    \left(\prod_{j = 1}^{k+l+m} \int_{\omega_{j}} \frac{E_{\alpha_{j}}(\omega_{j})}{i \omega_{j}}\right)
    \cr & \qquad \times
    \delta \left(\omega_{a} - \omega_{{k+1}} - \cdots - \omega_{{k+l}}\right)
    \delta(\omega - \omega_{a} + \omega_{c_{2k+1}} - \omega_{c_{1}})
    \left(\prod_{j = 0}^{k-1}
    \delta(\omega_{c_{2j+1}} - \omega_{c_{2j+2}} - \omega_{c_{2j+3}})\right)
    \cr & \qquad \times
    \delta(\omega_{c_{2}} - \omega_{{1}} - \omega_{{k+l+1}} - \cdots - \omega_{{k+l+m_{1}}})
    \cdots
    \delta(\omega_{c_{2k}} - \omega_{{k}} - \omega_{k+l+m - m_{k} + 1} - \cdots - \omega_{k+l+m})
    \cr & \qquad \times
    \sum_{\set{C}}
    \mathrm{Tr}\left[
    \hat{v}_{\mu\alpha_{k+1}\ldots \alpha_{k+l}}
    \hat{\mathcal{G}}^{C_{1}}(\omega_{c_{1}})
    \hat{v}_{\alpha_{1}\alpha_{k+l+1}\ldots \alpha_{k+l+m_{1}}}
    \hat{\mathcal{G}}^{C_{2}}(\omega_{c_{3}})
    \cdots
    \hat{\mathcal{G}}^{C_{k}}(\omega_{c_{2k-1}})
    \hat{v}_{\alpha_{k}\alpha_{k+l+m-m_{k}+1}\ldots \alpha_{k+l+m}}
    \hat{\mathcal{G}}^{C_{k+1}}(\omega_{c_{2k+1}})
    \right].
    \cr &
    \label{eq_j_klm}
  \end{align}
  The $n$-th order non-symmetrized conductivity tensor is given by the sum of the contributions satisfying $k + m + l = n$:
  \begin{multline}
    \tilde{\sigma}_{\mu; \alpha_{1}\ldots \alpha_{n}}(\omega_{1}, \ldots, \omega_{n})
    =
    -\frac{e^{n+1}}{V} \left(\prod_{j=1}^{n}\frac{1}{i \omega_{j}}\right) \sum_{klm}^{k+l+m=n} \sum_{m_{1}+\cdots+m_{k}}^{=m} \frac{1}{l!} \frac{1}{(m_{1}+1)!} \cdots \frac{1}{(m_{k}+1)!}
    \\ \times
    \tilde{\chi}_{(\mu \alpha_{k+1}\ldots \alpha_{k+l}); (\alpha_{1} \alpha_{k+l+1} \ldots \alpha_{k+l+m_{1}}), \ldots, (\alpha_{k}\alpha_{n-m_{k}+1}\ldots\alpha_{n})}(\omega_{1}+\omega_{k+l+1} +
    \\ \cdots + \omega_{k+l+m_{1}}, \ldots, \omega_{k} + \omega_{k+l+\sum_{j=1}^{k-1} m_{j} + 1} + \cdots + \omega_{n}),
  \end{multline}
  where
  \begin{multline}
    \tilde{\chi}_{(\mu \alpha_{k+1}\ldots \alpha_{k+l}); (\alpha_{1} \alpha_{k+l+1} \ldots \alpha_{k+l+m_{1}}), \ldots, (\alpha_{k}\alpha_{n-m_{k}+1}\ldots\alpha_{n})}(\omega_{1}+\omega_{k+l+1} + \cdots
    \\ + \omega_{k+l+m_{1}}, \ldots, \omega_{k} + \omega_{k+l+\sum_{j=1}^{k-1} m_{j} + 1} + \cdots + \omega_{n})
    \cr
    =
    -i \frac{1}{\hbar^{k}}
    \int_{\omega_{c}}
    \sum_{\set{C}} \mathrm{Tr}\left[ \hat{v}^{\mu\alpha_{k+1}\ldots \alpha_{k+l}}
    \left\{
    \hat{\mathcal{G}}^{C_{1}}(\omega_{c}) \hat{v}_{\alpha_{1}\alpha_{k+l+1}\ldots \alpha_{k+l+m_{1}}}\hat{\mathcal{G}}^{C_{2}} (\omega_{c} - \omega_{1} - \omega_{k+l+1} - \cdots - \omega_{k+l+ m_{1}})
    \right. \right.\cr
    \qquad \qquad
    \times \cdots \times \left.\left.
    \hat{v}_{\alpha_{k}\alpha_{n-m_{k}+1}\ldots \alpha_{n}} \hat{\mathcal{G}}^{C_{k+1}}(\omega_{c} - \omega_{[k]} - \omega_{k+l+1} - \cdots - \omega_{k+l+m}) \right\} \right].
  \end{multline}
  The $n$-th ($n > 1$) order conductivity tensor is obtained by symmetrizing the non-symmetrized ones $\tilde{\sigma}$ as
  \begin{align}
    \sigma_{\mu; \alpha_{1}\ldots \alpha_{n}}(\omega_{1}, \ldots, \omega_{n})
    =
    \frac{1}{n!} \sum_{\set{P}} \tilde{\sigma}_{\mu; \alpha_{p_{1}}\ldots \alpha_{p_{n}}}(\omega_{p_{1}}, \ldots, \omega_{p_{n}}),
    \label{eq_cond_n_sym}
  \end{align}
  where $\sum_{\set{P}}$ represents the sum over all permutations of $(1,2,\cdots,n)$.
  The explicit forms of $\sigma_{\mu; \alpha_{1}\ldots \alpha_{n}}(\omega_{1}, \ldots, \omega_{n})$ up to $n = 3$ are given by
  \begin{align}
     & \sigma_{\mu; \alpha}(\omega)
    =
    \frac{ie^{2}}{V (\omega + i\gamma)}
    \left\{v_{\mu\alpha}^{(0)} + \chi_{\mu; \alpha}(\omega)\right\},
    \label{eq_cond_1_}
  \end{align}
  \begin{multline}
    \sigma_{\mu; \alpha, \beta}(\omega_{1}, \omega_{2})
    =
    \frac{e^{3}}{2V (\omega_{1} + i \gamma) (\omega_{2} + i \gamma)}
    \biggl\{
    \frac{1}{2} v_{\mu\alpha\beta}^{(0)}
    +
    \frac{1}{2}
    \tilde{\chi}_{\mu; (\alpha\beta)}(\omega_{1} + \omega_{2})
    +
    \tilde{\chi}_{(\mu\alpha); \beta}(\omega_{2})
    +
    \tilde{\chi}_{\mu; \alpha, \beta}(\omega_{1}, \omega_{2})
    \biggr\} \\
    + [(\alpha, \omega_{1}) \leftrightarrow (\beta, \omega_{2})],
    \label{eq_cond_2_}
  \end{multline}
  and
  \begin{multline}
    \sigma_{\mu; \alpha_{1}, \alpha_{2}, \alpha_{3}} (\omega_{1}, \omega_{2}, \omega_{3})
    =
    -\frac{i e^{4}}{6V (\omega_{1} + i \gamma) (\omega_{2} + i \gamma) (\omega_{3} + i \gamma)}
    \\
    \times\biggl\{
    \frac{1}{6} v_{\mu\alpha_{1}\alpha_{2}\alpha_{3}}^{(0)}
    + \frac{1}{6}
    \tilde{\chi}_{\mu; (\alpha_{1} \alpha_{2} \alpha_{3})}(\omega_{1}+\omega_{2}+\omega_{3})
    + \frac{1}{2}
    \tilde{\chi}_{(\mu \alpha_{1}); (\alpha_{2} \alpha_{3})}(\omega_{2}+\omega_{3})
    + \frac{1}{2}
    \tilde{\chi}_{(\mu \alpha_{2} \alpha_{3}); \alpha_{1}}(\omega_{1})
    \\
    + \frac{1}{2} \tilde{\chi}_{\mu; \alpha_{1}, (\alpha_{2} \alpha_{3})}(\omega_{1}, \omega_{2}+\omega_{3})
    + \frac{1}{2} \tilde{\chi}_{\mu; (\alpha_{1} \alpha_{2}), \alpha_{3}}(\omega_{1}+\omega_{2}, \omega_{3})
    +\tilde{\chi}_{(\mu \alpha_{3});\alpha_{1}, \alpha_{2}}(\omega_{1}, \omega_{2})
    + \tilde{\chi}_{\mu; \alpha_{1},\alpha_{2},\alpha_{3}}(\omega_{1}, \omega_{2}, \omega_{3})
    \biggr\}
    \\
    + [(\alpha_{1}, \omega_{1}) \leftrightarrow (\alpha_{2}, \omega_{2})]
    + [(\alpha_{2}, \omega_{2}) \leftrightarrow (\alpha_{3}, \omega_{3})]
    + [(\alpha_{3}, \omega_{3}) \leftrightarrow (\alpha_{1}, \omega_{1})]
    \label{eq_cond_3_}.
  \end{multline}
  Note that the broadening factor $\gamma$ is introduced by replacing $\omega_{j}$ with $\omega_{j} + i \gamma$.
\end{widetext}

\section{Derivation of linear conductivity in the velocity gauge}
\label{sec:lcon_vel}

Here, we give the derivation of the linear conductivity in the velocity gauge, Eqs.~(\ref{eq_cond_1_drude}) and (\ref{eq_cond_1_bc}).
For notational simplicity, $\bm{k}$ is omitted in $O_{nm}(\bm{k})$, $\epsilon_{n\bm{k}}$, and so on.

$v_{\mu\alpha}^{(0)}$ and $\chi_{\mu; \alpha} (\omega)$ in Eq.~(\ref{eq_cond_1_}) are explicitly given by
\begin{align}
   & v_{\mu\alpha}^{(0)}
  =
  \sum_{\bm{k} n} f_{n} v_{\mu\alpha}^{nn},
  \label{eq_v0}                    \\
   & \chi_{\mu; \alpha} (\omega) =
  \sum_{\bm{k} n \neq m}
  \frac{f_{nm}}{\epsilon_{nm} + \hbar (\omega + i\gamma)}  v_{\mu}^{nm} v_{\alpha}^{mn},
  \label{eq_chi_vv}
\end{align}
where $\epsilon_{nm} \equiv \epsilon_{n} - \epsilon_{m}$ and $f_{nm} \equiv f_{n} - f_{m}$.
Using Eq.~(\ref{eq_velo_1}), $v_{\mu\alpha}^{nn}$ in Eq.~(\ref{eq_v0}) is written as
\begin{align}
  v_{\mu\alpha}^{nn} =
  \frac{1}{\hbar^{2}} \partial_{\mu} \partial_{\alpha} \epsilon_{n}
  - \sum_{n \neq m} \frac{v_{\mu}^{nm} v_{\alpha}^{mn} + v_{\alpha}^{nm} v_{\mu}^{nm}}{\epsilon_{nm}}.
  \label{aeq_v2d}
\end{align}
Then, Eq.~(\ref{eq_v0}) is reexpressed as
\begin{align}
  v_{\mu\alpha}^{(0)}
   & =
  \frac{1}{\hbar^{2}} \sum_{\bm{k} n}
  f_{n} \partial_{\mu} \partial_{\alpha} \epsilon_{n}
  -
  \sum_{\bm{k} n \neq m}
  \frac{f_{nm}}{\epsilon_{nm}} v_{\mu}^{nm}v_{\alpha}^{mn}.
  \label{eq_v0_}
\end{align}
Using Eq.~(\ref{eq_chi_vv}) and (\ref{eq_v0_}), the linear conductivity is eventually expressed as
\begin{multline}
  \sigma_{\mu; \alpha}(\omega)
  =
  \frac{i e^{2}}{V(\omega + i \gamma)}
  \sum_{\bm{k} n}
  f_{n} \partial_{\mu} \partial_{\alpha} \epsilon_{n} \cr
  +
  \frac{\hbar e^{2}}{iV}
  \sum_{\bm{k} n \neq m}
  \frac{f_{nm}}{\epsilon_{nm}}
  \frac{1}{\epsilon_{nm} + \hbar (\omega + i\gamma)}
  v_{\mu}^{nm}v_{\alpha}^{mn}.
  \label{eq_cond_om}
\end{multline}

Taking the static limit in Eq.~(\ref{eq_cond_om}), we obtain
\begin{align}
  \sigma_{\mu; \alpha}
   & =
  \frac{e^{2}}{\gamma \hbar^{2} V}
  \sum_{\bm{k}} \sum_{a}
  f_{n} \partial_{\mu} \partial_{\alpha} \epsilon_{n}
  -\frac{e^{2}}{\hbar V}
  \sum_{\bm{k} n}
  \epsilon_{\mu\alpha\beta}
  f_{n}
  \Omega^{\beta}_{n}.
  \label{aeq_cond_1_ab_3}
\end{align}
The first and second terms correspond to the Drude term, Eq.~(\ref{eq_cond_1_drude}) and BC term, Eq.~(\ref{eq_cond_1_bc}), respectively.

\bibliographystyle{apsrev4-2.bst}

%

\end{document}